\documentclass[%
reprint,
superscriptaddress,
aps,
prab,
]{revtex4-2}

\usepackage{graphicx}
\usepackage{dcolumn}
\usepackage{bm}
\usepackage{amsmath,amssymb,}

\usepackage[utf8]{inputenc}
\usepackage[T1]{fontenc}
\usepackage{booktabs, array, mathptmx, float, tabularx, booktabs, lipsum, multirow}
\usepackage{siunitx, xcolor}
\usepackage[version=4]{mhchem}
\graphicspath{{./figs/}} %
\usepackage[colorlinks,linkcolor=blue,anchorcolor=blue,citecolor=blue]{hyperref}
\usepackage{graphicx}
\usepackage{subcaption}
\usepackage{makecell}

\begin{document}

\preprint{APS/123-QED}

\title{Performance and tolerance study of the rectilinear cooling channel for a muon collider}

\author{Ruihu Zhu}
\affiliation{
	Institute of Modern Physics, Chinese Academy of Sciences, Lanzhou 730000, China
}
\affiliation{
	University of Chinese Academy of Sciences, Beijing 100049, China
}

\author{Chris Rogers}
\email{chris.rogers@stfc.ac.uk}
\affiliation{
	STFC Rutherford Appleton Laboratory, Didcot OX11 0QX, United Kingdom
}

\author{Jiancheng Yang}
\email{yangjch@impcas.ac.cn}
\affiliation{
	Institute of Modern Physics, Chinese Academy of Sciences, Lanzhou 730000, China
}
\affiliation{
	University of Chinese Academy of Sciences, Beijing 100049, China
}

\author{He Zhao}
\affiliation{
	Institute of Modern Physics, Chinese Academy of Sciences, Lanzhou 730000, China
}
\affiliation{
	University of Chinese Academy of Sciences, Beijing 100049, China
}

\author{Cheng Guo}
\affiliation{
	Institute of Modern Physics, Chinese Academy of Sciences, Lanzhou 730000, China
}
\affiliation{
	University of Chinese Academy of Sciences, Beijing 100049, China
}

\author{Jiangdong Li}
\affiliation{
	Institute of Modern Physics, Chinese Academy of Sciences, Lanzhou 730000, China
}
\affiliation{
	University of Chinese Academy of Sciences, Beijing 100049, China
}





\date{\today}

\begin{abstract}
The muon collider has the potential to be a powerful tool for the exploration of frontiers in particle physics. In order to reach high luminosity, the 6D emittance of the muon beam needs to be reduced by several orders of magnitude. The cooling process for a muon collider involves two parts; initial six-dimensional cooling and final transverse cooling. This paper focuses on the former and proposes a conceptual design of the rectilinear cooling channel with additional dipole magnets. In this paper, we first introduce a general method for designing the rectilinear cooling channel. Subsequently, we apply this method to develop two rectilinear cooling channels before and after a bunch merging system. Furthermore, we investigate the impact on cooling performance by employing $\pi$-mode RF cavities and considering the effect of errors in the magnetic and RF fields.

\end{abstract}

\maketitle


\section{Introduction}
Particle physicists seek to collide particles at the highest possible energy with high luminosity. Historically, this has been achieved using electron-positron colliders, which additionally offer clean events enabling precision studies. However, achieving multi-TeV collision energies is challenging for electron colliders due to the small mass of the electron, resulting in significant energy loss from synchrotron radiation \cite{book_1}. Hadron colliders have also been used. Synchrotron radiation is suppressed in hadron colliders, owing to the large proton mass, but the wide proton parton distribution leads to a reduction in the energy of collision products relative to the centre of mass energy. Muons, on the other hand, have a much larger mass compared to electrons, making them less affected by synchrotron radiation. Additionally, muons have electron-like properties, making muon colliders a promising choice for high-energy physics \cite{Towards_a_Muon_Collider}.

A technical challenge for the muon collider arises from the large emittance of the initial muon beam. This emittance significantly exceeds the acceptance limits of downstream accelerator components and is unsuitable for achieving high luminosity collisions, requiring a dedicated cooling channel to reduce the beam emittance. Additionally, due to the extremely short lifetime of muons (\textasciitilde2$\mu$s in the rest frame), the cooling process must be completed before the muons decay completely. This requirement makes ionization cooling the only feasible method to cool the muons. Ionization cooling can be classified into two types: 4D cooling and 6D cooling. In 4D ionization cooling, when the muon beam passes through an absorber, it simultaneously loses transverse and longitudinal momentum, with the longitudinal momentum restored by RF cavities. Consequently, the transverse phase space of the muons decreases over time. 4D ionization cooling was demonstrated by the Muon Ionization Cooling Experiment (MICE) collaboration \cite{MICE1,MICE2}. This arrangement does not achieve reduction of longitudinal emittance. In order to realize 6D ionization cooling, a dipole field and wedge-shaped absorber is envisaged to be introduced into the apparatus \cite{NEUFFER200426}. This setup ensures that particles with higher longitudinal momentum traverse a thicker part of the wedge, leading to greater longitudinal momentum loss. Consequently, both longitudinal and transverse emittance can be reduced simultaneously.

Four main types of ionization cooling channel have been developed in the past. The first one is a ring-shaped cooling channel which uses tilted solenoids to generate the dispersion and bend the beam \cite{Ionization_Ring}. Simulations indicate that it successfully reduces the 6D emittance of the muons by a factor of \textasciitilde50. However, significant challenge remains with injection into and extraction from the ring. To address this challenge, another cooling channel, known as the Guggenheim design \cite{Guggenheim}, was proposed. Simulation results indicate that it achieves nearly the same cooling performance as the cooling ring. However, as the cooling cells in Guggenheim are set on a vertical helix, it will be very difficult to construct. The third one is the helical cooling channel. This is also a 6D cooling channel, but it uses helical magnets and homogeneous absorbers instead of solenoids and wedge absorbers \cite{helical_cooling}. The fourth one is rectilinear cooling channel. In this design, the components of the cooling cell are the same as those used in the cooling ring or Guggenheim. However, in the rectilinear design, the cooling cells are arranged along a straight line. The rectilinear channel, initially proposed by Balbekov \cite{Balbekov}, has a much simpler geometry compared to the Guggenheim design. This simplicity makes it easier to construct. Additionally, unlike the fixed focusing and dispersion in the cooling ring, the rectilinear channel allows for adjustment of these parameters at different stages. This flexibility enables the rectilinear channel to reduce the emittance of muons to much smaller values \cite{Tapered_channel,MAP}. For these reasons, we choose the rectilinear cooling channel as the baseline in our studies. 

For the design of the rectilinear cooling segments before and after the bunch merging system in this paper, uncoupled RF cells are used. These RF cells are short and have a small RF phase difference between adjacent cells. Although shorter RF cells have a higher transit time factor, leading to more acceleration for a given electric field (see details in Section \ref{section_4}), each uncoupled RF cell requires a cryo-module feed-through, which complicates the engineering process. The International Muon Collider Collaboration (IMCC) proposes a 6D muon cooling demonstrator \cite{demonstrator} using $\pi$-mode RF \cite{pi_mode_RF}, where the accelerating phase difference between two adjacent RF cells is $\pi$. Sites such as CERN in Switzerland, Fermi National Laboratory in the US and the High Intensity heavy-ion Accelerator Facility \cite{HIAF, CNUF} in China are under consideration. $\pi$-mode RF offers numerous advantages, such as its compact waveguide structure and the requirement for only one RF coupler and cryo-module feed-through to supply all the RF cells. However, it also has some disadvantages, including high RF power requirements for each coupler and a low transit time factor. Since a low transit time factor might negatively impact beam dynamics, it is crucial to examine the impact on cooling performance with $\pi$-mode RF. 

A tolerance study including magnetic and RF errors is also performed to assess the robustness of the cooling lattice. It is important to note that this tolerance study is conducted in only one cooling stage with $\pi$-mode RF for convenience, but the results are expected to hold for designs with uncoupled RF cells as well.

This paper is structured as follows: Section \ref{section_2} provides a review of the theory of 6D ionization cooling. Section \ref{section_3} presents the parameters and simulation results of the proposed rectilinear cooling channel. In Section \ref{section_4}, we discuss the simulation results using $\pi$-mode RF and analyze how magnetic and RF errors influence the cooling performance. Finally, Section \ref{section_5} summarizes the conclusions drawn from this study.
\section{Principles of 6D ionization cooling}\label{section_2}
When it undergoes ionization cooling, a muon beam gradually loses both transverse and longitudinal momentum owing to the ionization of atoms within the absorber material. RF cavities restore longitudinal momentum but not transverse momentum. As a result, the muons’ momentum becomes more parallel, leading to a reduction in emittance. The evolution of transverse emittance is described as follows \cite{NEUFFER200426}:
\begin{equation}\label{evolution_without_longitudinal}
    \frac{d\varepsilon_T}{ds}=-\frac{1}{\beta^2}\frac{dE_\mu}{ds}\frac{\varepsilon_T}{E_\mu}+\frac{1}{\beta^3}\frac{\beta_{T}E_s^{2}}{2E_{\mu}m_{\mu}c^{2}L_R}
\end{equation}
where $\varepsilon_T$ is the normalized transverse emittance, $E_\mu$ is the muon beam energy, $m_\mu$ is the muon mass, $\beta$ is the muon particle velocity, $c$ is the speed of light, $\beta_T$ is the transverse beta value, $dE_\mu/ds$ is the energy loss per unit length, $L_R$ is the radiation length of absorber material and $E_s$ is the characteristic scattering energy (\textasciitilde13.6 MeV). 

The energy loss $dE_\mu/ds$ can be estimated by the Bethe-Bloch equation \cite{ParticleDataGroup:1998hll}:
\begin{equation}\label{energy_loss}
\frac{dE_\mu}{ds}=4{\pi}N_Ar_e^2m_ec^2{\rho}\frac{Z}{A}\left[\frac{1}{\beta^2}\ln\left(K\gamma^2\beta^2\right)-1-\frac{\delta}{2\beta^2}\right]
\end{equation}
where $r_e$ is the classical electron radius, $\rho$ is the density, $N_A$ is Avogadro's number, $A$ is the atomic weight, $Z$ is the atomic number and $m_e$ is the electron mass. $K=2m_ec^2/I$ and $I$ is the mean excitation energy. $\delta$ is the density effect factor which is negligible for the muons with longitudinal momentum being around 200 MeV.

The first term of Eq. (\ref{evolution_without_longitudinal}) can be interpreted as the cooling from energy loss due to atomic ionization, while the second term represents the heating from the Coulomb scattering. The equilibrium transverse emittance is defined when $d\varepsilon_T/ds$ in Eq.(\ref{evolution_without_longitudinal}) is 0 \cite{NEUFFER200426}:
\begin{equation}\label{eq_emittance_without_longitudinal}
\varepsilon_{T,eq}=\frac{\beta_{T}E_s^{2}}{2{\beta}m_{\mu}c^2L_R\left|\frac{dE_\mu}{ds}\right|}
\end{equation}

From Eq.~(\ref{eq_emittance_without_longitudinal}), it is evident that in order to achieve a lower equilibrium transverse emittance, two key factors should be considered. Firstly, the focusing at the absorber should be strong, indicated by a smaller transverse beta value. Secondly, the absorber material should possess a large product of $L_R$ and $\left|\frac{dE_\mu}{ds}\right|$, which is satisfied by materials with low atomic numbers, such as liquid hydrogen and lithium hydride.

For a muon collider, it is crucial to also decrease the longitudinal emittance to meet the required acceptance of downstream accelerator components. To achieve longitudinal cooling, a well-known scheme called emittance exchange is employed. This involves using wedge-shaped absorbers and introducing dispersion. The introduction of dispersion causes the beam to spread transversely, allowing particles with higher momentum to pass through a thicker part of the absorber and lose more energy, thereby reducing the longitudinal emittance at the cost of increasing the transverse emittance. The value of dispersion must be carefully chosen to simultaneously achieve both transverse and longitudinal cooling, also known as 6D cooling. 

For the 6D cooling using the wedge, formulas for the evolution of  transverse and longitudinal emittance have been provided in \cite{NEUFFER200426,MAP,Neuffer_2017} and are shown as below:

\begin{equation}\label{evolution_transverse_emittance}
\frac{d\varepsilon_T}{ds}=-\frac{g_T}{\beta^2E_\mu}\frac{dE_\mu}{ds}\varepsilon_T+\frac{\beta_TE_s^2}{2\beta^3m{_\mu}c^2L_RE_\mu}
\end{equation}
\begin{equation}\label{evolution_longitudinal_emittance}
\frac{d\varepsilon_L}{ds}=-\frac{g_L}{\beta^2E_\mu}\frac{dE_\mu}{ds}\varepsilon_L+\frac{\beta_L}{2}\frac{d\left<\Delta{E^2}\right>}{ds}
\end{equation}
where $g_T$ and $g_L$ are the transverse and longitudinal partition numbers, respectively, and $\beta_L$ is the longitudinal beta function. They can be expressed as follows \cite{Neuffer_2017}:
\begin{equation}\label{transverse_partition_number}
g_T=1-\frac{D}{w}
\end{equation}
\begin{equation}\label{longitudinal_partition_number}
g_L=\frac{2\gamma^2-2\ln[K(\gamma^2-1)]}{\gamma^2\ln[K(\gamma^2-1)]-(\gamma^2-1)}+\frac{D}{w}
\end{equation}
\begin{equation}\label{longitudinal_beta}
\beta_L=\sqrt{\frac{\lambda_{RF}\beta^3{\gamma}m_{\mu}c^2\alpha_p}{2\pi{eV}^{'}\cos{\varphi_s}}}
\end{equation}
where $D$ is the dispersion, $w$ is the distance between the beam center and the apex of the wedge, $\alpha_p$ is the slip factor, which can be estimated as $-1/\gamma^2$ ($\gamma$ being the Lorentz factor) since it is an approximation in the linac and the rectilinear cooling channel is roughly a linac, $V^{'}$ is the average RF gradient, $\varphi_s$ is the RF phase and $\lambda_{\rm{RF}}$ is the RF wavelength.

The partition numbers $g_T$ and $g_L$ describe how the total damping from cooling is distributed between the transverse and longitudinal planes. The longitudinal beta function $\beta_L$ represents the focusing strength in the longitudinal plane.

The second term in Eq.~(\ref{evolution_longitudinal_emittance}) arises from random fluctuations in the energy loss known as energy straggling. It can be approximately described as \cite{energy_straggling}:
\begin{equation}\label{energy_straggling}
\frac{d\left<\Delta{E^2}\right>}{ds}=4{\pi}(r_e{\gamma}m_ec^2)^2n_e\left(1-\frac{\beta^2}{2}\right)
\end{equation}

The equillibrium emittance can be derived when $d\varepsilon{/}ds=0$, so, from Eq.~(\ref{evolution_transverse_emittance}) and Eq.~(\ref{evolution_longitudinal_emittance}), we get \cite{Neuffer_2017}:

\begin{equation}\label{eq_transverse_emittance}
\varepsilon_T^{eq}=\frac{\beta_{T}E_s^{2}}{2\left|\frac{dE_\mu}{ds}\right|{\beta}g_{T}m_{\mu}c^{2}L_R}
\end{equation}
\begin{equation}\label{eq_longitudinal_emittance}
\varepsilon_L^{eq}=\frac{\beta_{L}m_ec^2\gamma^2(1-\frac{\beta^2}{2})}{2g_L\beta{m_\mu}c^2[\frac{\ln(K\gamma^2\beta^2)}{\beta^2}-1]}
\end{equation}

The expressions for the evolution of transverse and longitudinal emittance can also be obtained from Eq.~(\ref{evolution_transverse_emittance}) and Eq.~(\ref{evolution_longitudinal_emittance}) \cite{Neuffer_2017}:
\begin{equation}\label{emittance_evolution}
\varepsilon_i(s)=(\varepsilon_{i,0}-\varepsilon_{i,eq})\exp\left({-\frac{s}{L_{cool,i}}}\right)+\varepsilon_{i,eq}
\end{equation}
where $i=T~or~L$ corresponding to the transverse or longitudinal direction, $\varepsilon_{i,0}$ is the initial emittance and $L_{cool,i}$ is the cooling length shown in Eq.~(\ref{cooling_length}).
\begin{equation}\label{cooling_length}
L_{cool,i}=\left(\frac{g_i}{\beta^2E_\mu}\left<\frac{dE_\mu}{ds}\right>\right)^{-1}
\end{equation}
where $dE_\mu/ds$ with angular brackets is the energy loss averaging over the full transport length.

Eqs.~(\ref{eq_transverse_emittance}) to (\ref{cooling_length}) are used to calculate the theoretical emittance at the end of each rectilinear cooling stage discussed in section \ref{section_4}.
\section{Linear Lattice Optics and general design methods}\label{section_3}
\subsection{Lattice function and beam dynamics}
Studying the fundamental lattice features and beam dynamics in the rectilinear cooling channel is essential as it provides valuable guidance for the design and simulation of the channel, aiding in optimizing its performance and efficiency. We analyze the lattice function and beam dynamics of the channel without absorbers and RF.
\subsubsection{Transverse beta function, phase advance and momentum acceptance}
The beta function is a crucial parameter in accelerator physics as it represents the focusing strength of the external magnetic field. In the case of solenoids, due to cylindrical symmetry, it is more common to use the transverse beta function $\beta_T$ instead of individual beta function $\beta_x$ and $\beta_y$. The transverse beta function evolves as \cite{linear_optics}:
\begin{equation}\label{evolution_transverse_beta}
2\beta_T\beta_{T}^{''}-\beta_{T}^{'2}+4\beta_{T}^{2}k^2-4=0
\end{equation}
where $k$ is the solenoid focusing strength,
\begin{equation}\label{focusing_strength}
k=\frac{qB_{z}}{2p_z}
\end{equation}
where q is the charge of the particle, $B_z$ is the longitudinal magnetic field and $p_z$ is the longitudinal momentum.

One can solve Eq.~(\ref{evolution_transverse_beta}) periodically to obtain the value of transverse beta function along the cooling cell. 

Phase advance $\varphi$ is defined as:
\begin{equation}\label{phase_advance}
\varphi=\int\frac{1}{\beta_T}dz
\end{equation}
When the phase advance is close to or equal to $n\pi$ ($n=1,2,3,...$), the value of transverse beta function increases significantly leading to the integer or half-integer resonance.

It is evident that the phase advance of the cooling cell is dependent on the beam longitudinal momentum, as indicated by Eq.~(\ref{evolution_transverse_beta}), Eq.~(\ref{focusing_strength}) and Eq.~({\ref{phase_advance}). 
The momentum acceptance is defined as the range of beam longitudinal momentum values where the phase advance lies between $n\pi$ and $(n+1)\pi$.
It is crucial to ensure that the momentum acceptance remains sufficiently large (e.g., 5 or 6 times larger than the RMS longitudinal momentum spread) in order to minimize particle loss.

\subsubsection{Closed orbit and dispersion}
Given the symmetry of the magnetic field in the y direction around the middle point of the cooling cell (shown in Fig.~(\ref{field})), it is necessary for the closed orbits in both the x and y directions to exhibit symmetry as well. This symmetry requires that the derivatives of closed orbits at the boundary of the cooling cell are zero. Consequently, this simplifies the search for closed orbits, as the initial momenta in both the x and y directions can be set to zero. If the beam center is placed on the closed orbit, most particles in the beam will perform betatron oscillation around the closed orbit, resulting in the overall beam motion being nearly periodic and leading to reduced particle loss. The horizontal and vertical dispersion components can be calculated from the closed orbit as:
\begin{equation}\label{dispersion_x}
D_{x}=\frac{\Delta{x}}{\Delta{p}/p}
\end{equation} 
\begin{equation}\label{dispersion_y}
D_{y}=\frac{\Delta{y}}{\Delta{p}/p}
\end{equation} 
where $\Delta{x}$, $\Delta{y}$ and $\Delta{p}$ represent the difference of the closed orbit in x and y direction, respectively, and  momentum relative to the reference particle.
\begin{figure*}[htbp]
  \centering
  \begin{subfigure}[b]{6.45cm}
    \includegraphics[width=6.45cm]{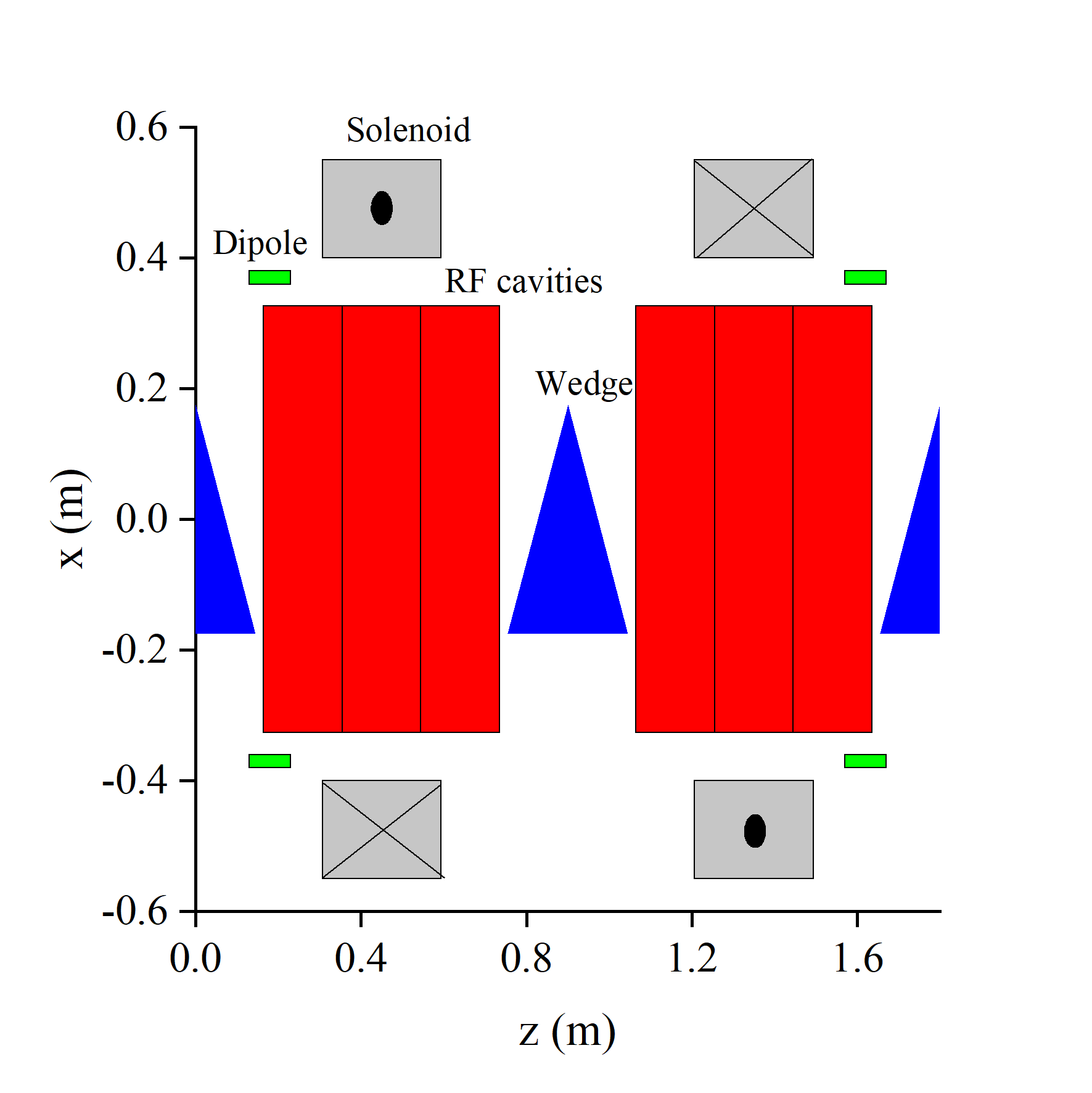}
    \caption{}
    \label{cell_layout_a}
  \end{subfigure}
  \begin{subfigure}[b]{6.45cm}
    \includegraphics[width=6.45cm]{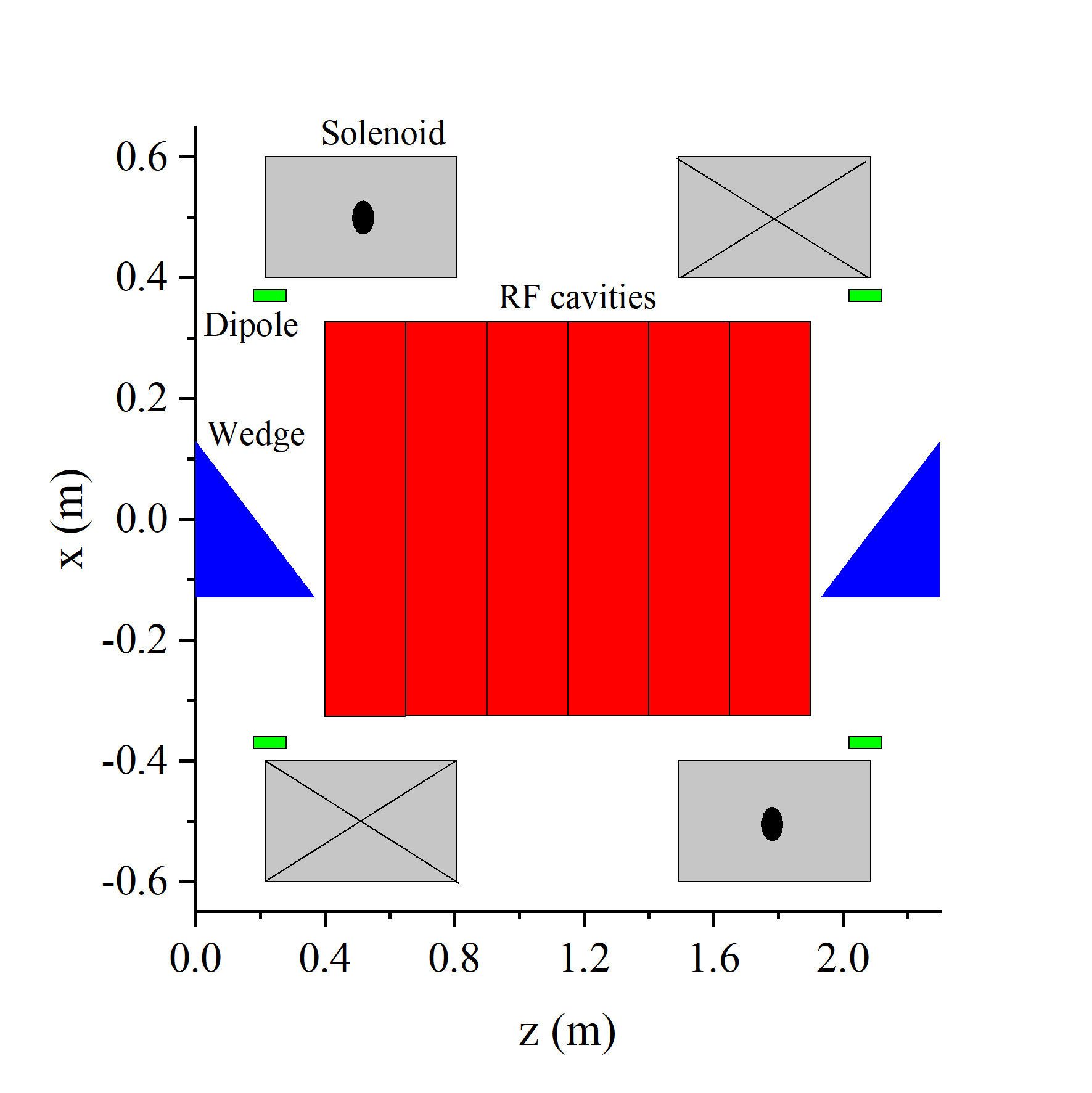}
    \caption{}
    \label{cell_layout_b}
  \end{subfigure}
  \caption{Schematics of two types of cooling cell layout: (a) A-type; (b) B-type}
  \label{cell_layout}
\end{figure*}
\subsection{Cooling channel design process}
The design process consists of four steps as follows:

a) Calculation of transverse beta function and momentum acceptance: The first step is to choose an appropriate transverse beta function value at the center of the wedge and a sufficient momentum acceptance. We use the well-known differential evolution algorithm \cite{scipy} to adjust the relative solenoid parameters (position, current density and length) and the target functions which need to be minimized are:
\begin{equation}\label{target_function_beta_before}
f=(\beta_T-\beta_{T,ref})^2+(\varphi_{low}-\pi)^2
\end{equation} 
\begin{equation}\label{target_function_beta_post}
f=(\beta_T-\beta_{T,ref})^2+(\varphi_{low}-2\pi)^2+(\varphi_{high}-\pi)^2
\end{equation} 
where $\beta_{T,ref}$ is the transverse beta function value we choose, $\varphi_{low}$ and $\varphi_{high}$ denote the phase advances of the cooling cell, which are obtained from the lowest and highest momenta of the chosen momentum acceptance. Eq.~(\ref{target_function_beta_before}) is for the phase advance of the cooling cell below $\pi$ and Eq.~(\ref{target_function_beta_post}) is for the phase advance between $\pi$ and 2$\pi$. 

b) Calculation of closed orbit and dispersion: The second step is to determine the value for the dispersion which decides the emittance exchange rate. As the dispersion is calculated from closed orbit difference shown in Eqs.~(\ref{dispersion_x}) and (\ref{dispersion_y}), we need to find the closed orbit first. The target function used to find the closed orbit is:
\begin{equation}\label{target_function_orbit}
f=(x_{final}-x_{init})^2+(y_{final}-y_{init})^2
\end{equation} 
The dispersion is controlled by the strength of the dipole field. It should be noted that the cell lattice used in this step is without the wedges and RF cavities.

c) Obtaining the list of RF parameters: The dispersion in the RF cavities region results in a coupled transverse and longitudinal beam motion. This means that finding the proper RF parameters involves more than just compensating for energy loss in the wedge absorber. It also requires maintaining the same closed orbit as in step b). We manually set the dispersion and choose the maximum accelerating gradient and accelerating phase as two variables. The wedge absorber length is adjusted manually in a certain range. The target function for finding the closed orbit in this step is the same as Eq.~(\ref{target_function_orbit}). Additionally, since selecting the correct longitudinal momentum of the reference particle affects the timing of the RF cavities, we iterate over a certain range of the reference particle’s z-momentum to obtain several lists of RF parameters. In this step, for each wedge absorber length, several lists of RF parameters corresponding to different reference momenta are obtained.

d) Running the multi-particle tracking simulation: The lists obtained from step c) are utilized as inputs for the RF cavities, and the multi-particle tracking simulation is initiated. G4Beamline-3.08 \cite{G4beamline} is employed to complete the tracking simulation, while the emittance calculation is carried out using the code Ecalc9f \cite{ecalc9f}. \verb|QGSP_BERT_EMX| is chosen for the physics model in G4Beamline for the tracking simulations, as it includes all relevant physics processes such as multiple scattering, energy straggling, energy loss, and muon decay. A 4$\sigma$ cut is applied in Ecalc9f for the emittance calculation. We introduce a merit factor to select the best outcome and quantify cooling efficiency. It is described as \cite{Towards_a_Muon_Collider}:
\begin{equation}\label{merit_factor}
M(s)=\frac{T(s)^2}{\frac{\varepsilon_T(s)}{\varepsilon_T(0)}\sqrt{\frac{\varepsilon_L(s)}{\varepsilon_L(0)}}}
\end{equation}
where T is the transmission, $\varepsilon_T(s)$ and $\varepsilon_T(0)$
 are the normalized transverse emittance at a specific position and start of the cooling section, respectively, while $\varepsilon_L(s)$ and $\varepsilon_L(0)$ refer to the normalized longitudinal emittance. This merit factor is indicative of the improvement in luminosity arising from the cooling provided by the rectilinear cooling channel.

All calculations and simulations in the above steps are performed in parallel using two AMD EPYC 7642 processors with a total of 96 cores.
\section{lattice parameters and tracking studies of the rectilinear cooling channel}\label{section_4}
\subsection{Layout of the cooling cell}
The basic lattice layout of one cell used in this paper is depicted in Fig.~(\ref{cell_layout}) which resembles the previous design \cite{MAP}. However, instead of tilting solenoids to generate dipole field, separate dipole magnets are incorporated to facilitate tuning of the dipole field, for example during commissioning.

As shown in Fig.~(\ref{cell_layout}), each cooling cell consists of solenoids with opposite polarity, dipole magnets for dispersion generation, RF cavities for beam energy loss compensation and liquid hydrogen ($\mathrm{LH_2}$) wedge absorbers. This paper utilizes two types of cooling cell layouts, referred to as A-type and B-type, which are shown in Fig.~(\ref{cell_layout_a}) and Fig.~(\ref{cell_layout_b}). The primary difference between these layouts lies in the period of the squared longitudinal magnetic field, $B_z^2$. For the A-type layout, the period is half the length of the cooling cell, whereas for the B-type layout, the period is the full length of the cooling cell. 

The A-type lattice has the advantage that the transverse beta function at the center is equal to that at the start and end of the cooling cell, enabling the placement of a wedge absorber at the middle of the cell. Assuming fixed energy loss in wedge absorbers, this arrangement results in a reduction of the length of each wedge absorber, consequently lowering the average transverse beta function as described by Eq.~(\ref{average_beta}). 
\begin{equation}\label{average_beta}
\beta_{T,ave}=\frac{\int_0^{L}\beta_T \, dz}{L}
\end{equation} 
where L is the length of the absorber.

The B-type lattice can achieve a smaller transverse beta function compared to the A-type, which aids in further emittance reduction. However, the momentum acceptance of the B-type layout is smaller than that of the A-type. 

Windows are also taken into account for both the liquid hydrogen absorbers and RF cavities, although they are not displayed in Fig.~(\ref{cell_layout}). Beryllium (Be) is chosen as the window material for the absorber due to its low atomic number, which has minimal impact on the cooling performance. Although there is some risk associated with using beryllium in conjunction with liquid hydrogen, an R\&D program is currently underway to establish safe design parameters for beryllium.  Beryllium is also selected as the window material for the RF because it can increase the operational gradient \cite{rf_cavity}. For the absorber, the window thickness varies from 300 $\mu$m (stage 1 in pre-merging section) to 40 $\mu$m (stage 10 in post-merging section).  For the RF, the window thickness varies from 120 $\mu$m (stage 1 in pre-merging section) to 10 $\mu$m (stage 10 in post-merging section). The geometry of the absorber and RF windows are both simple disks. Each absorber and RF cavity is enclosed with two windows at both ends. The windows of the wedge absorbers are oriented to completely cover the sides of the triangular prism-shaped wedges.

In characterizing the fringe field of the dipole magnets, we employ the expressions derived from \cite{fringe_field}. The fringe field components are described by the following equations:
\begin{equation}\label{dipole_fringe_y}
B_y=B_0\frac{1+e^{a_1z}\cos{a_1y}}{1+2e^{a_1z}\cos{a_1y}+e^{2a_1z}}
\end{equation}
\begin{equation}\label{dipole_fringe_z}
B_z=B_0\frac{-e^{a_1z}\sin{a_1y}}{1+2e^{a_1z}\cos{a_1y}+e^{2a_1z}}
\end{equation}
where $B_0$ is the nominal dipole field strength, $a_1$ is a coefficient set to 5 in the simulations, and z and y are the coordinates normalized by the dipole magnet aperture. It is also noteworthy that while only the integrated dipole field significantly influences the closed orbit of the particles motion, utilizing fully Maxwellian fringe field expressions is always more physical. 
\begin{table*}
\caption{\label{cell_parameters}Main parameters for the cooling cells in each stage. A and B denote the pre-merging and post-merging section respectively. Liquid hydrogen is used as the wedge absorber material for all stages.}
\begin{ruledtabular}
\begin{tabular}{ccccccccccccccc}
\textbf{ } &\makecell[c]{Cell \\length \\(m)} & \makecell[c]{Stage \\length \\(m)}& \makecell[c]{Pipe \\radius \\(cm)} & \makecell[c]{Max. \\ on-axis \\ $\mathrm{B_z}$ (T)}  & \makecell[c]{Integrated\\ $\mathrm{B_y}$ (T·m)}& \makecell[c]{Transverse \\beta \\(cm)}& \makecell[c]{Dispersion \\(mm)}& \makecell[c]{On-axis \\wedge \\length \\(cm)}& \makecell[c]{Wedge \\apex \\angle \\(deg)}& \makecell[c]{RF \\frequency \\(MHz)}& \makecell[c]{Number \\of \\RF cells} & \makecell[c]{RF cell\\length \\(cm)} & \makecell[c]{Max. \\RF \\gradient\\ (MV/m)} & \makecell[c]{RF \\phase \\(deg)}\\ \hline
A-Stage 1 & 1.8 & 104.4 & 28 & 2.5  &0.102& 70 & -60& 14.5 & 45 & 352 & 6 & 19 & 27.4& 18.5 \\
 A-Stage 2& 1.2& 106.8& 16& 3.7 &0.147& 45& -57& 10.5& 60& 352& 4& 19& 26.4&23.2\\
 A-Stage 3& 0.8& 64.8& 10& 5.7 &0.154& 30& -40& 15& 100& 704& 5& 9.5& 31.5&23.7\\
 A-Stage 4& 0.7& 86.8& 8& 7.2 &0.186& 23& -30& 6.5& 70& 704& 4& 9.5& 31.7&25.7\\
 B-Stage 1& 2.3& 50.6& 23& 3.1 &0.106& 35& -51.8& 37& 110& 352& 6& 25& 21.2&29.9\\
 B-Stage 2& 1.8& 66.6& 19& 3.9 &0.138& 30& -52.4& 28& 120& 352& 5& 22& 21.7& 27.2\\
 B-Stage 3& 1.4& 84& 12.5& 5.1 &0.144& 20& -40.6& 24& 115& 352& 4& 19& 24.9& 29.8\\
 B-Stage 4& 1.2& 66& 9.5& 6.6 &0.163& 15& -35.1& 20& 110& 352& 3& 22& 24.3& 31.3\\
 B-Stage 5& 0.8& 44& 6& 9.1 &0.116& 10& -17.7& 12.5& 120& 704& 5& 9.5& 22.5& 24.3\\
 B-Stage 6& 0.7& 38.5& 4.5& 11.5 &0.0868& 6& -10.6& 11& 130& 704& 4& 9.5& 28& 22.1\\
 B-Stage 7& 0.7& 28& 3.75& 13 &0.0882& 5& -9.8& 10& 130& 704& 4& 9.5& 28.5& 18.4\\
 B-Stage 8& 0.65& 46.15& 2.85& 15.8 &0.0726& 3.8& -7.0& 7& 140& 704& 4& 9.5& 27.1& 14.5\\
 B-Stage 9& 0.65& 33.8& 2.3& 16.6 &0.0694& 3& -6.1& 7.5& 140& 704& 4& 9.5& 29.7& 11.9\\
 B-Stage 10& 0.63& 29.61& 2& 17.2 &0.0691& 2.7& -5.7& 6.8& 140& 704& 4& 9.5& 24.9& 12.2\\
\end{tabular}
\end{ruledtabular}
\end{table*}
\subsection{Design of the pre-merging cooling section}
We use the output beam file from the front end of the previously proposed Neutrino factory \cite{neutrino_factory} as the starting point for our tracking simulation, following a similar approach to the previous design \cite{MAP}. However, we have made changes by selecting 352 MHz and 704 MHz as our RF frequencies instead of the 325 MHz and 650 MHz used in the previous design. To match the new 352 MHz RF, we adjust the input beam by compressing it in time by a factor of 325/352 and stretching its z-momentum by 352/325. This ensures the input beam matches the new 352 MHz RF frequency while keeping its longitudinal phase space volume conserved.
\begin{figure*}[htbp]
  \centering
  \begin{subfigure}[b]{6.45cm}
    \includegraphics[width=6.45cm]{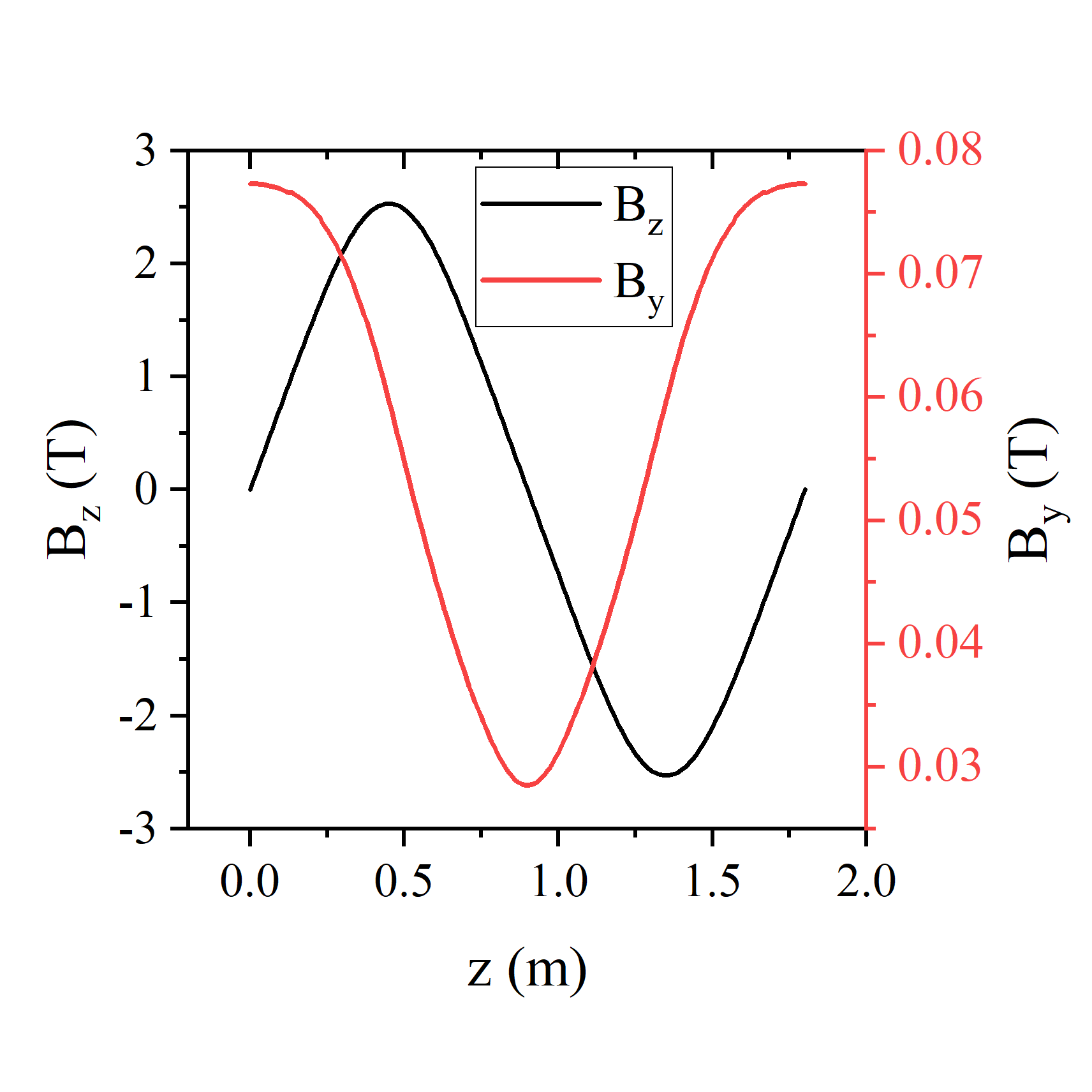}
    \caption{}
    \label{field_a}
  \end{subfigure}
  \begin{subfigure}[b]{6.45cm}
    \includegraphics[width=6.45cm]{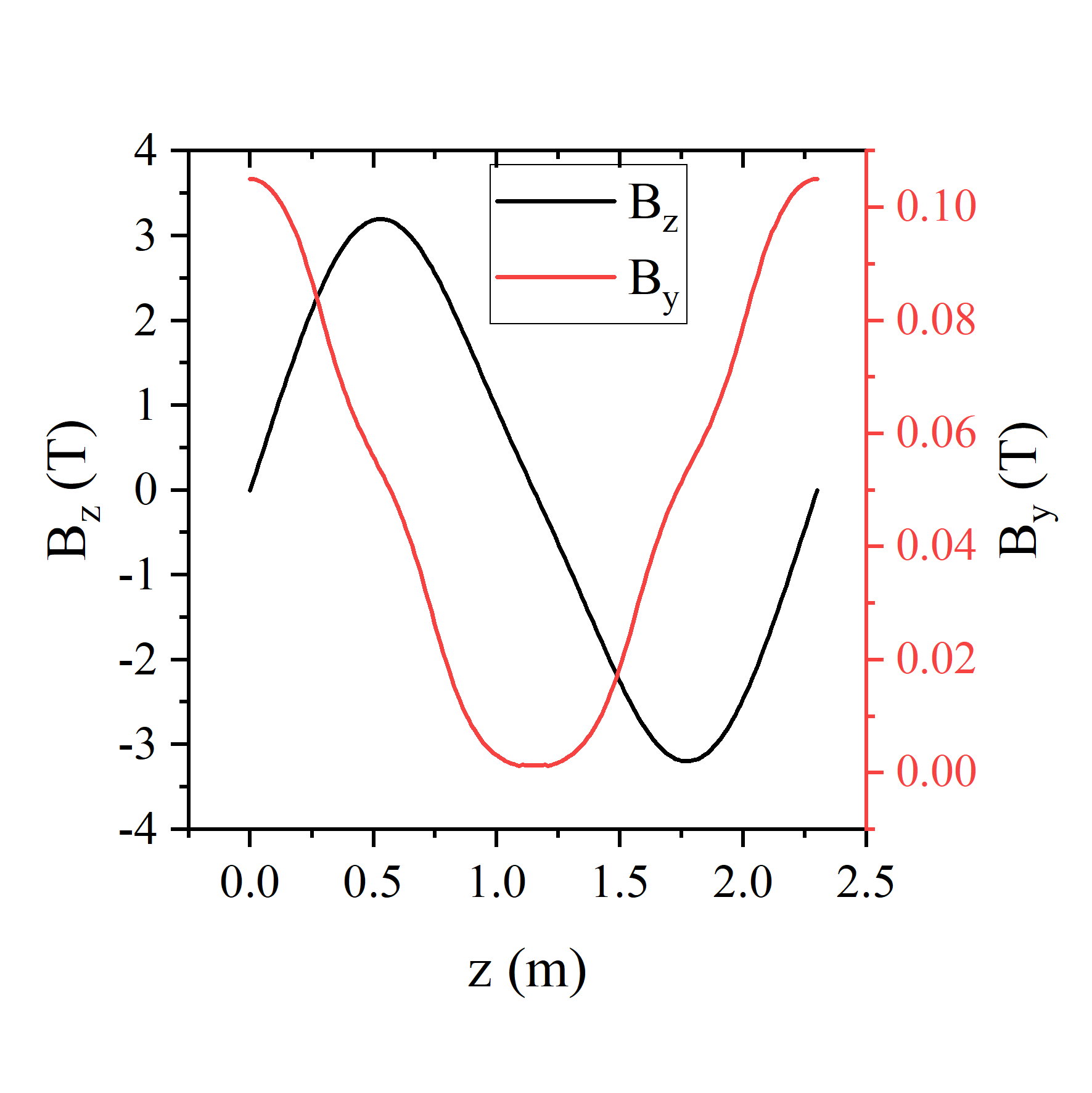}
    \caption{}
    \label{field_b}
  \end{subfigure}
  \caption{On-axis $\mathrm{B_z}$ and $\mathrm{B_y}$ of the cooling cell: (a) stage 1 in pre-merging section; (b) stage 1 in post-merging section}
  \label{field}
\end{figure*}
\begin{figure*}[htbp]
  \centering
  \begin{subfigure}[b]{6.45cm}
    \includegraphics[width=6.45cm]{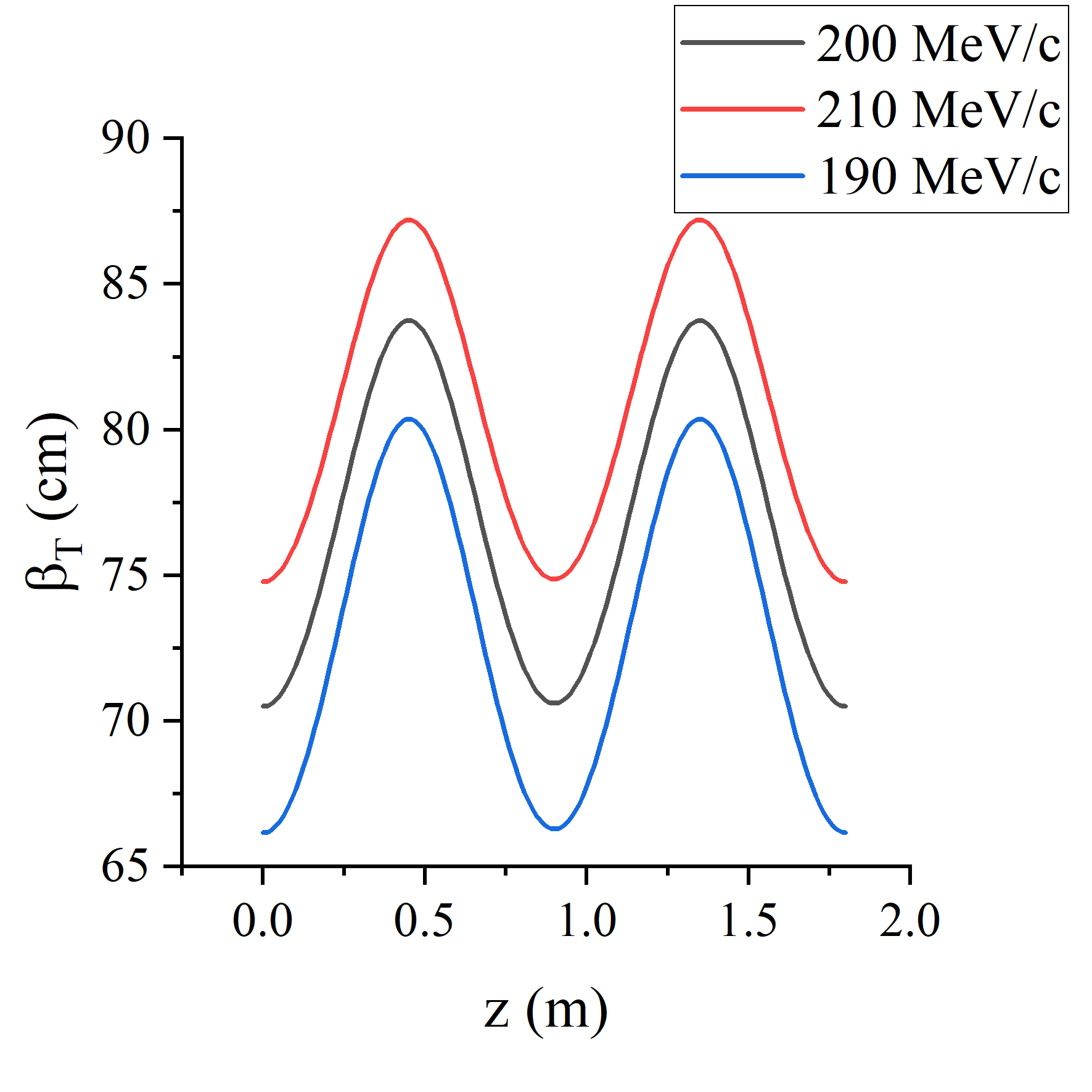}
    \caption{}
    \label{beta_z_stage1_before}
  \end{subfigure}
  \begin{subfigure}[b]{6.45cm}
    \includegraphics[width=6.45cm]{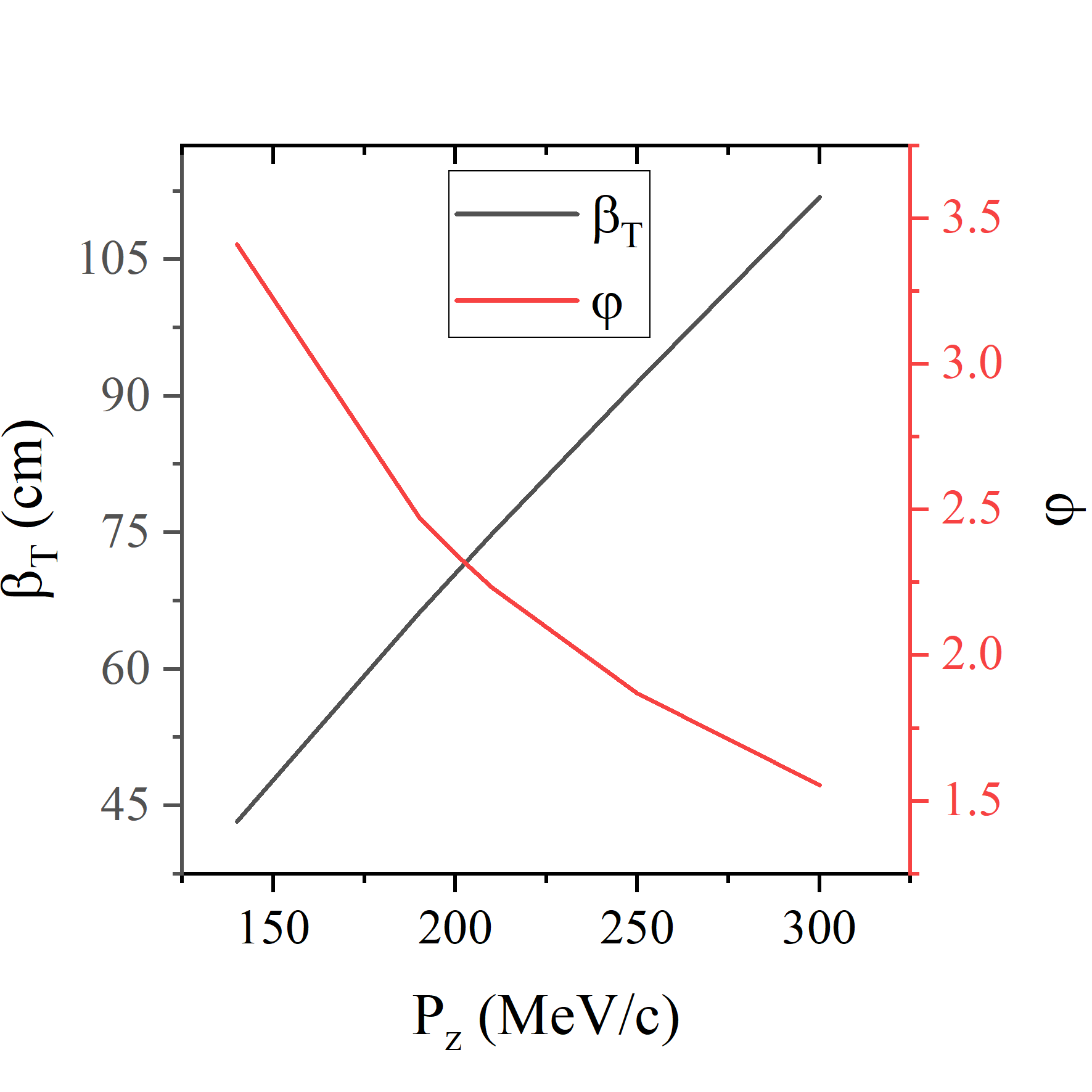}
    \caption{}
    \label{beta_pz_stage1_before}
  \end{subfigure}
  \begin{subfigure}[b]{6.45cm}
    \includegraphics[width=6.45cm]{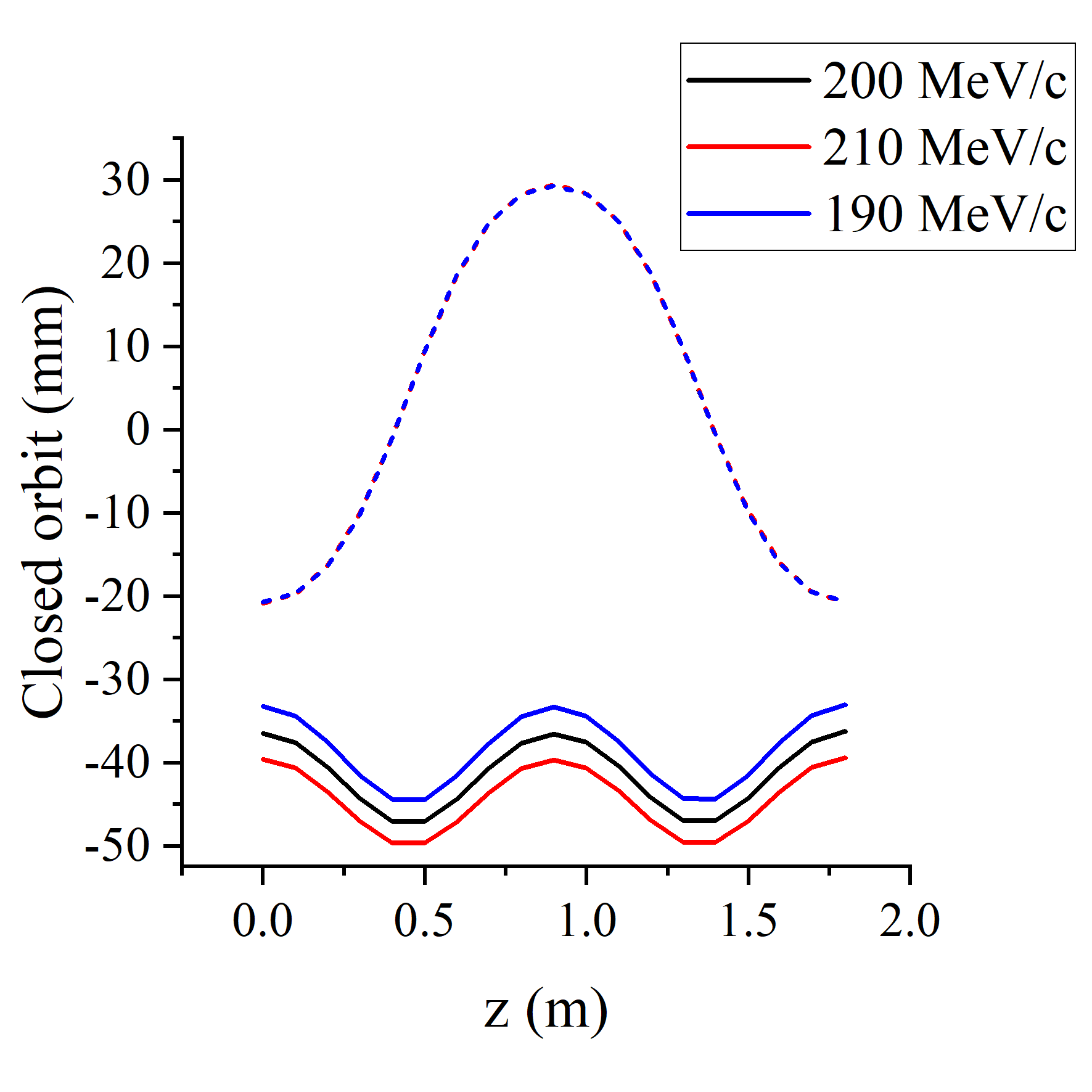}
    \caption{}
    \label{closed_orbit_stage1_before}
  \end{subfigure}
  \begin{subfigure}[b]{6.45cm}
    \includegraphics[width=6.45cm]{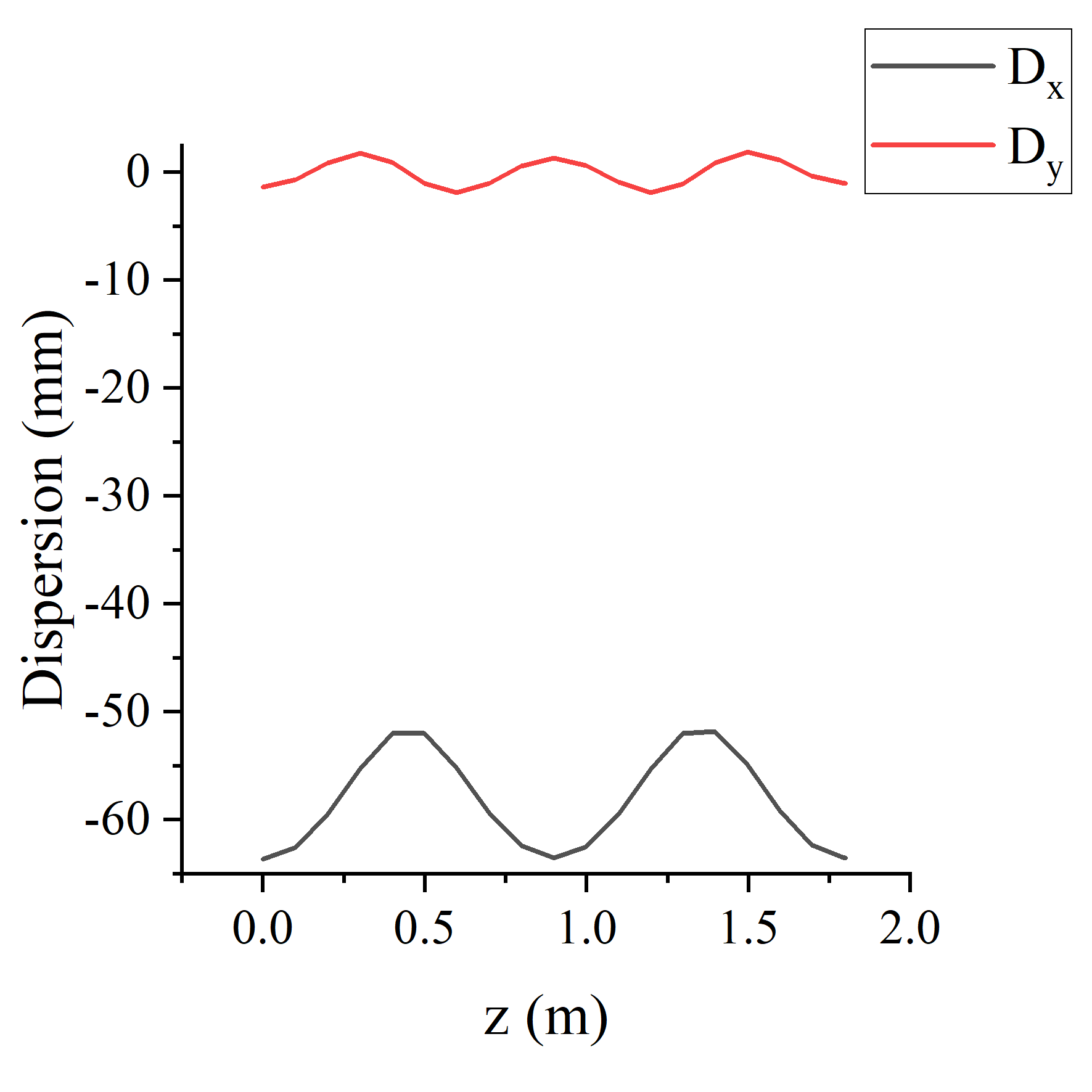}
    \caption{}
    \label{dispersion_stage1_before}
  \end{subfigure}
  \caption{Evolution of lattice parameters of the cooling cell in stage 1 before bunch merging: (a) transverse beta function versus position for 3 different z-momenta; (b) transverse beta function at the wedge absorber and phase advance of the cooling cell versus z-momentum; (c) closed orbit versus position for 3 different z-momenta (solid line: x plane, dashed line: y plane); (d) dispersion versus position (calculated at z-momentum of 200 MeV/c)}
  \label{lattice_parameters_evolution_before}
\end{figure*}
\begin{figure*}[htbp]
  \centering
  \begin{subfigure}[b]{6.45cm}
    \includegraphics[width=6.45cm]{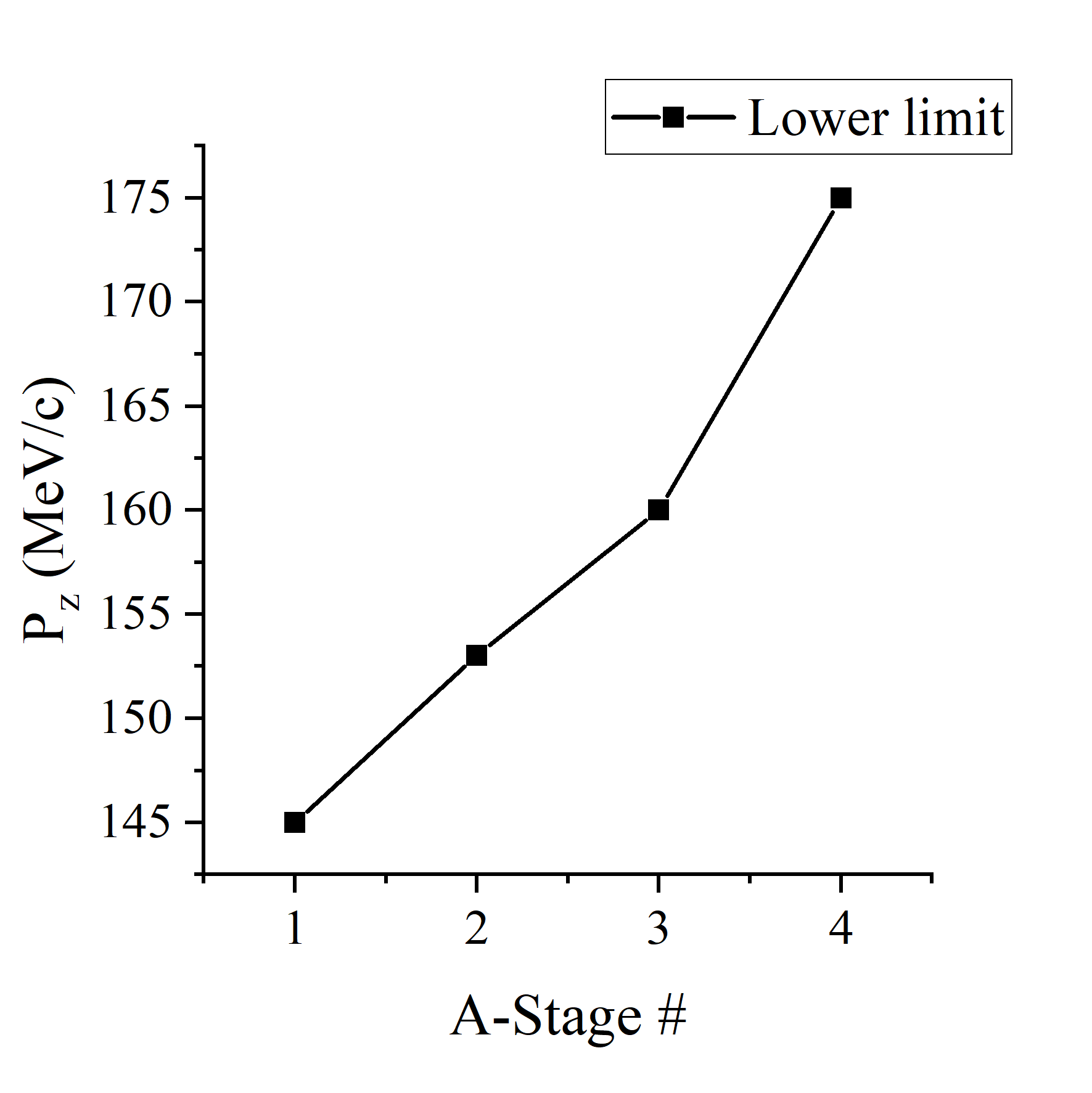}
    \caption{}
    \label{momentum_acceptance_before}
  \end{subfigure}
  \begin{subfigure}[b]{6.45cm}
    \includegraphics[width=6.45cm]{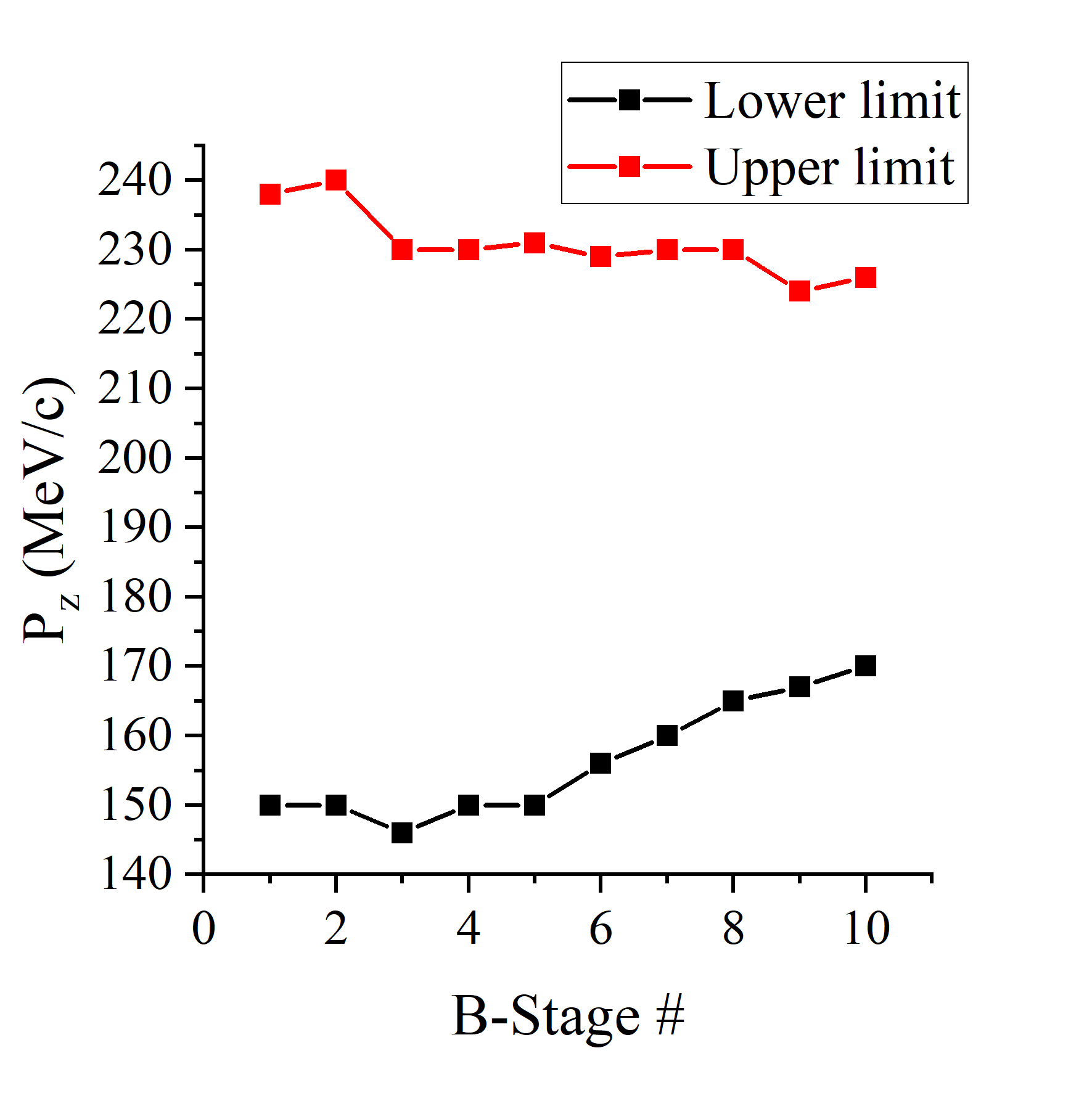}
    \caption{}
    \label{momentum_acceptance_post}
  \end{subfigure}
  \caption{Lower and upper limit of the momentum acceptance of the cooling cell in each stage: (a) acceptance in pre-merging section; (b) acceptance in post-merging section}
\label{fig_momentum_acceptance}
\end{figure*}
\begin{figure*}[htbp]
  \centering
  \begin{subfigure}[b]{6.45cm}
    \includegraphics[width=6.45cm]{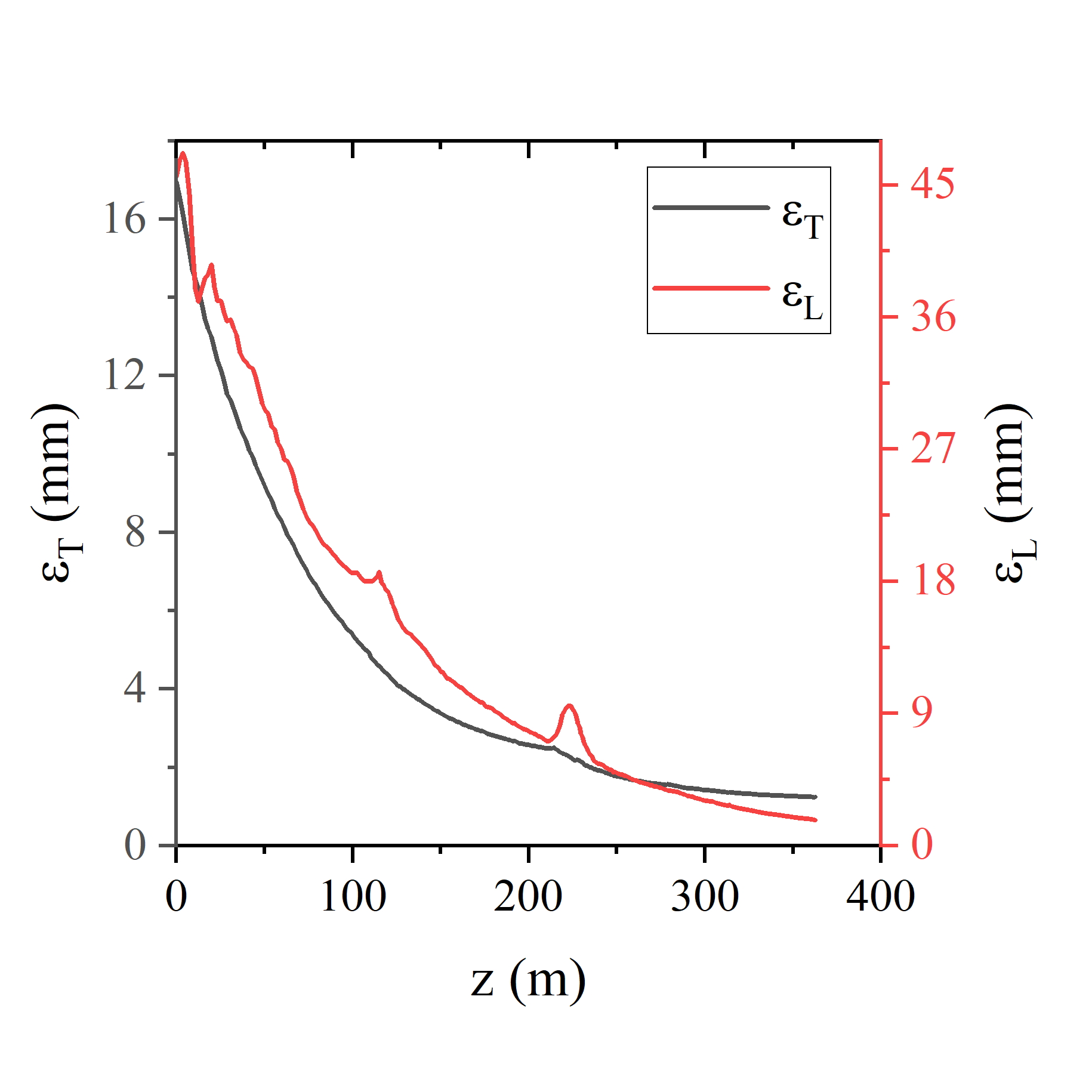}
    \caption{}
    \label{emittance_before}
  \end{subfigure}
  \begin{subfigure}[b]{6.45cm}
    \includegraphics[width=6.45cm]{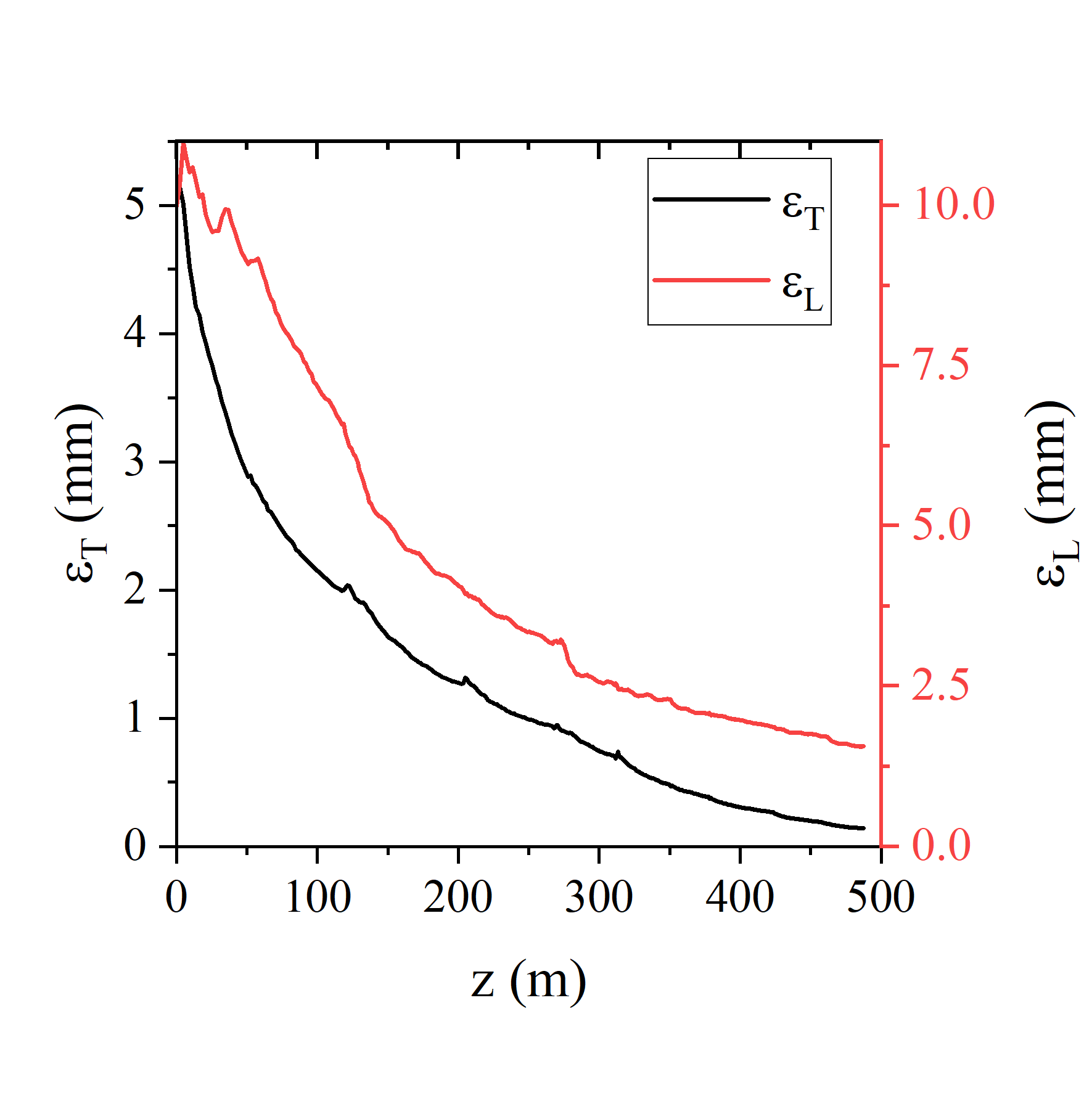}
    \caption{}
    \label{emittance_post}
  \end{subfigure}
  \caption{Evolution of the normalized transverse and longitudinal emittance: (a) emittance evolution in pre-merging section; (b) emittance evolution in post-merging section}
\label{fig_emittance_evolution}
\end{figure*}
\begin{figure*}[htbp]
  \centering
  \begin{subfigure}[b]{6.45cm}
    \includegraphics[width=6.45cm]{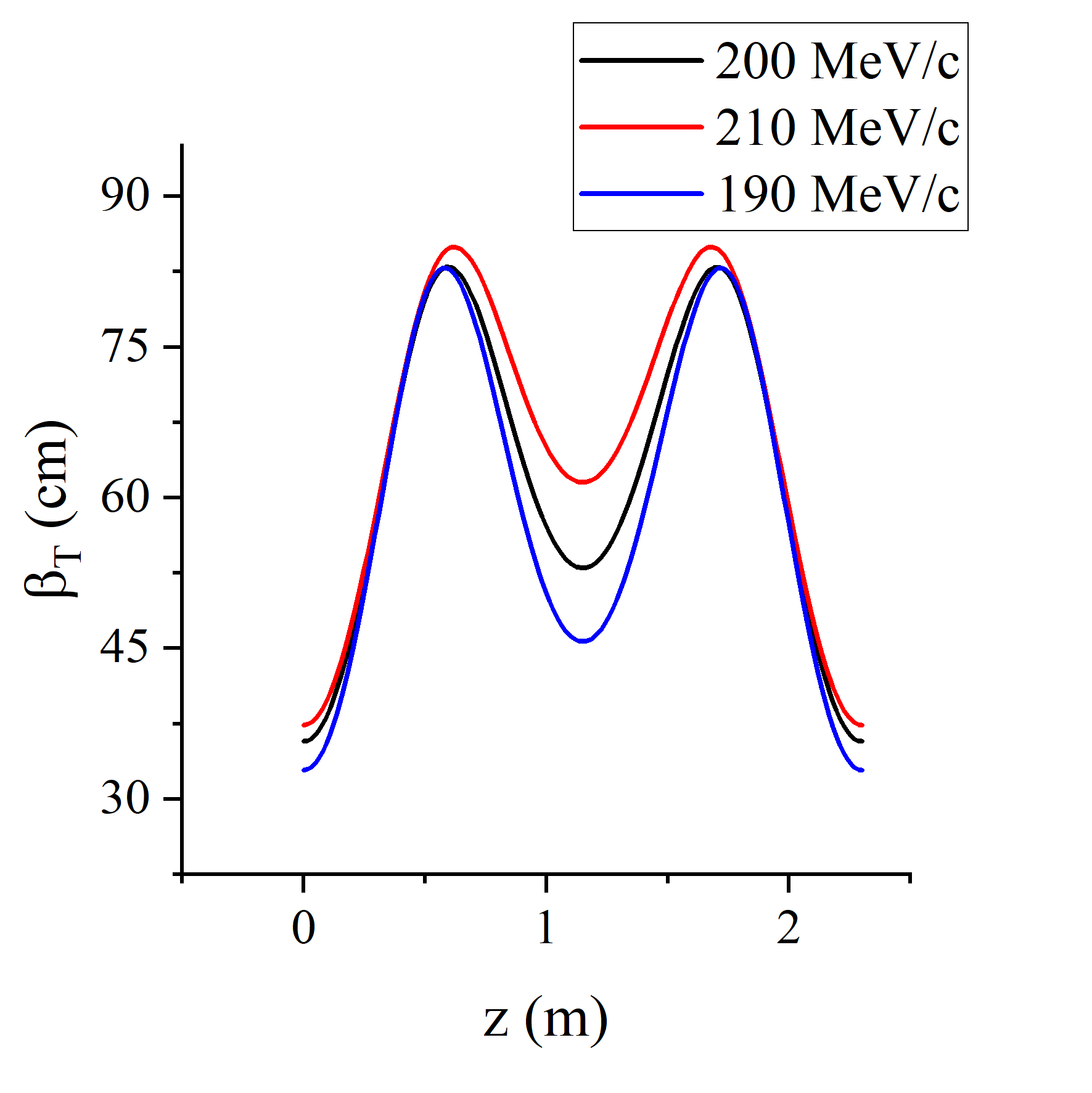}
    \caption{}
    \label{beta_z_stage1_post}
  \end{subfigure}
  \begin{subfigure}[b]{6.45cm}
    \includegraphics[width=6.45cm]{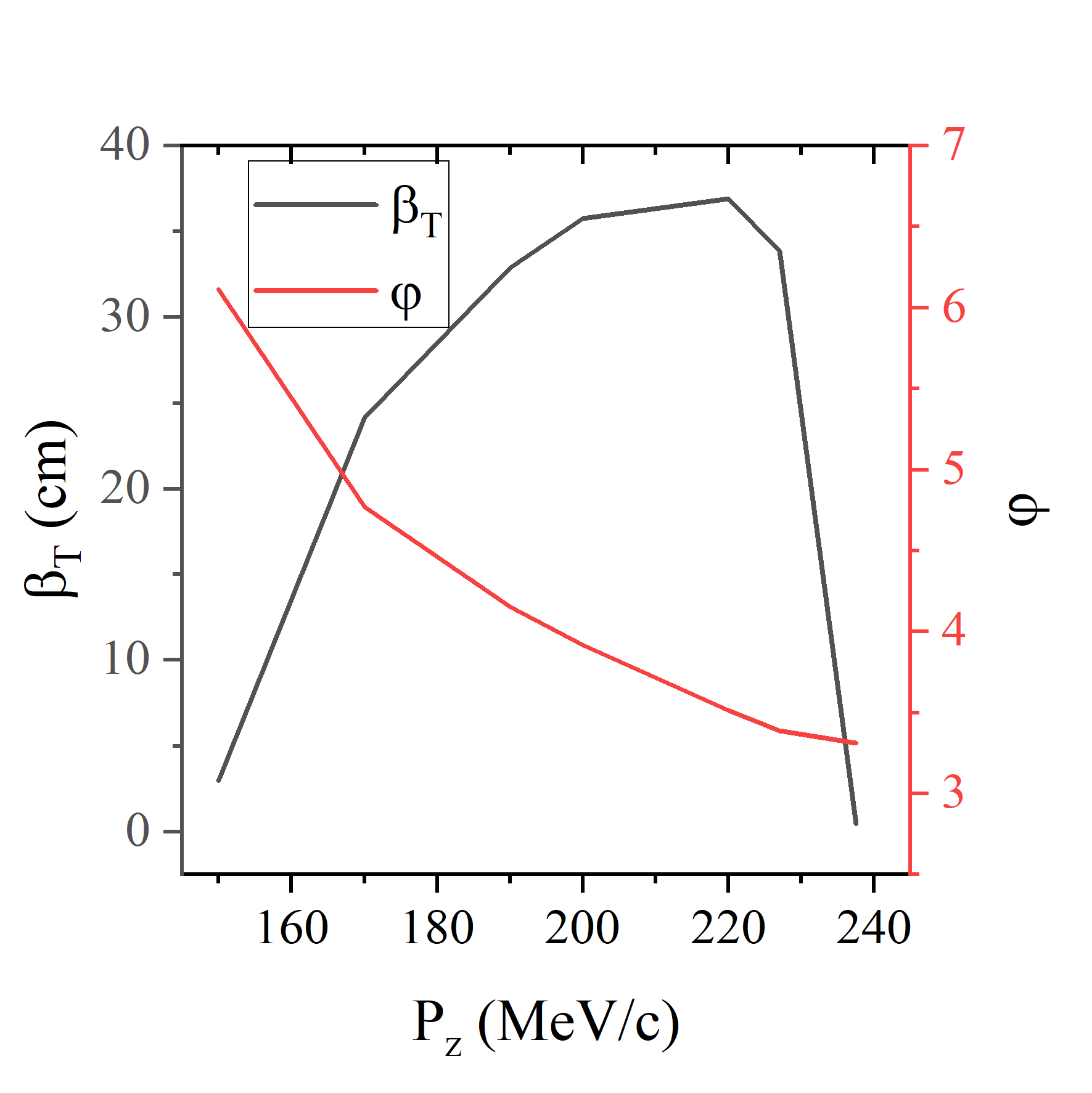}
    \caption{}
    \label{beta_pz_stage1_post}
  \end{subfigure}
  \begin{subfigure}[b]{6.45cm}
    \includegraphics[width=6.45cm]{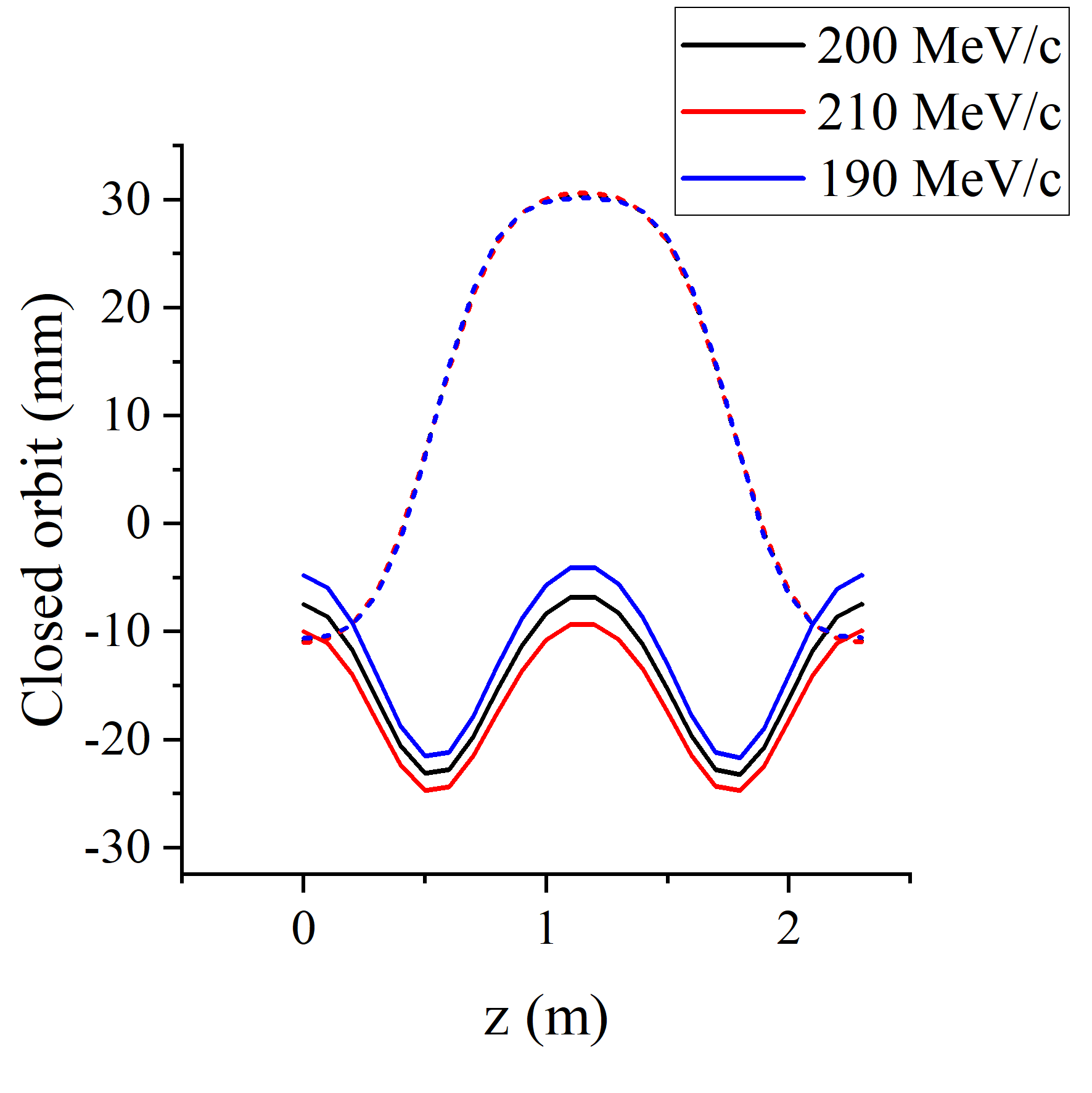}
    \caption{}
    \label{closed_orbit_stage1_post}
  \end{subfigure}
  \begin{subfigure}[b]{6.45cm}
    \includegraphics[width=6.45cm]{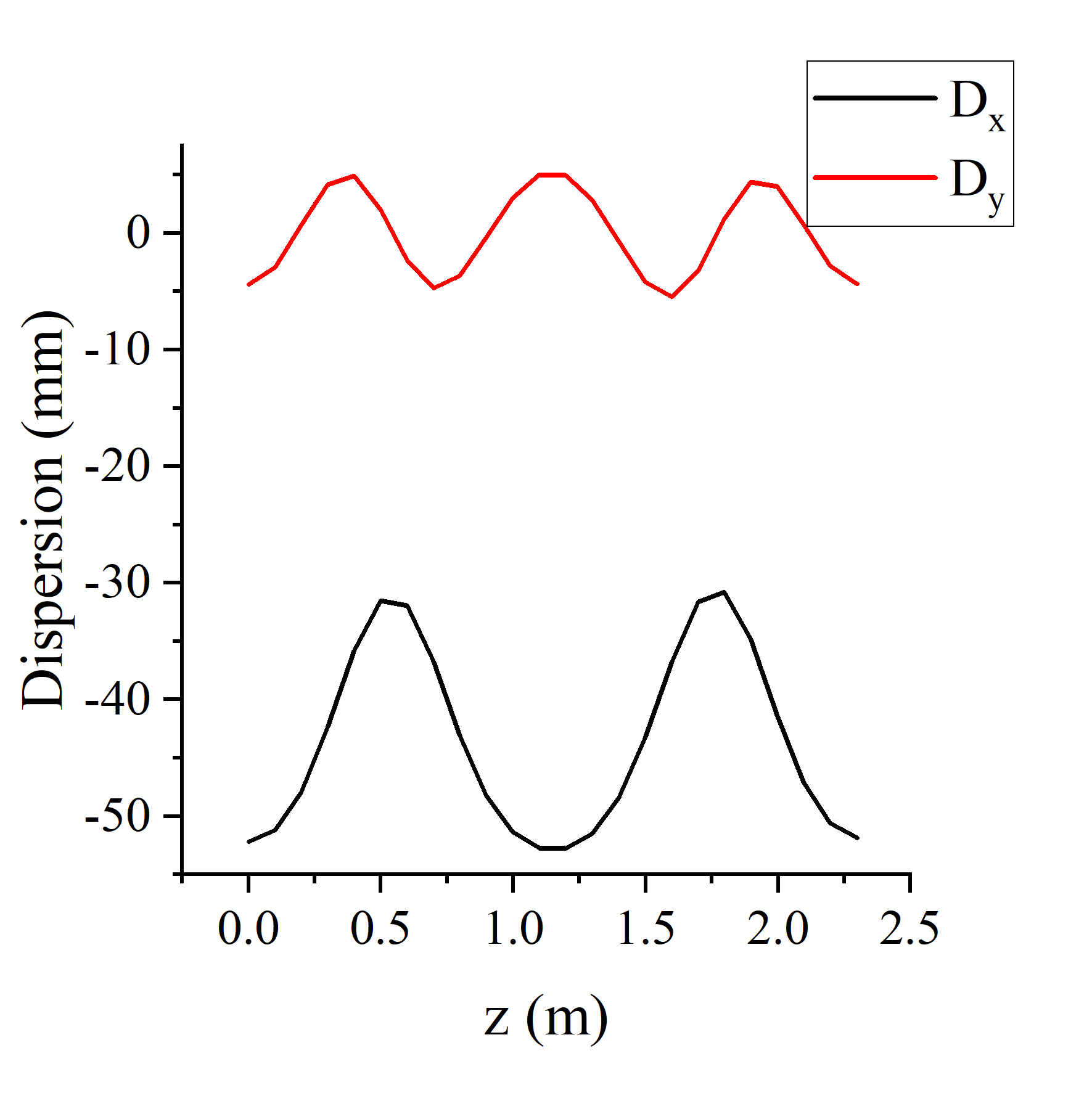}
    \caption{}
    \label{dispersion_stage1_post}
  \end{subfigure}
  \caption{Evolution of lattice parameters of the cooling cell in stage 1 after bunch merging: (a) transverse beta function versus position for 3 different z-momenta; (b) transverse beta function at the wedge absorber and phase advance of the cooling cell versus z-momentum; (c) closed orbit versus position for 3 different z-momenta (solid line: x plane, dashed line: y plane); (d) dispersion versus position (calculated at z-momentum of 200 MeV/c)}
\label{lattice_parameters_evolution_post}
\end{figure*}
\begin{table*}
\caption{\label{emittance_results_before}Simulation results on the normalized emittance and transmission at the end of each stage berfore merging}
\begin{ruledtabular}
\begin{tabular}{cccccc}
\textbf{} & $\varepsilon_{\rm{T,sim}}$ (mm) & $\varepsilon_{\rm{L,sim}}$ (mm) & $\varepsilon_{\rm{6D,sim}}$ (mm$^3$) & Transmission & Reference momentum (MeV/c)\\
\hline
Start& 16.96& 45.53& 13500& &\\
A-Stage 1& 5.165& 18.31& 492.6&75.2\% &253\\
A-Stage 2& 2.473& 7.113& 44.03&84.4\% &211\\
A-Stage 3& 1.556& 3.880& 9.594&85.6\% &211\\
A-Stage 4& 1.239& 1.741& 2.861&91.3\% &219\\ 
\end{tabular}
\end{ruledtabular}
\end{table*}

A 4-stage rectilinear channel is utilized to cool the beam to meet the required initial emittance for the bunch merging system. Given the large emittance of the input beam, it is necessary to avoid over-focusing in the first stage. Therefore, a relatively large beta function value of 70 cm is chosen at the wedge. The layout of the cooling cell in stage 1 is depicted in Fig.~(\ref{cell_layout_a}), including two solenoids with opposite polarity, two positive dipole magnets, six 352 MHz RF cells, and two liquid hydrogen wedge absorbers. Detailed parameter information is provided in Table~\ref{cell_parameters}. Fig.~(\ref{field_a}) illustrates the on-axis $\mathrm{B_z}$ generated from the solenoid coils in G4Beamline and the $\mathrm{B_y}$ generated from Eqs.~(\ref{dipole_fringe_y}) and (\ref{dipole_fringe_z}) of the cooling cell in stage 1 in the pre-merging section. The shape of the on-axis $\mathrm{B_z}$ is sinusoidal and flips at the middle of the cell to eliminate angular momentum accumulation. The maximum on-axis $\mathrm{B_z}$ is 2.5 T, corresponding to a large beta function value compared to later cooling cells. As shown in Fig.~(\ref{beta_z_stage1_before}), the beta functions at the start, middle, and end of the cooling cell are the same, with wedge absorbers placed at these three positions. Fig.~(\ref{beta_pz_stage1_before}) shows the dependence of the transverse beta function at the wedge absorber and the phase advance of the cooling cell on momentum. The transverse beta function at the wedge absorber is approximately proportional to the z-momentum. The phase advance at 145 MeV/c exceeds $\mathrm{\pi}$, indicating a momentum acceptance above ~145 MeV/c. It can be seen from Fig. (\ref{momentum_acceptance_before}) that the momentum acceptance gradually decreases throughout these four stages. Fig.~(\ref{closed_orbit_stage1_before}) shows how the closed orbit changes in the x and y directions for momenta of 200 MeV/c, 210 MeV/c, and 190 MeV/c. As expected, only the x-direction orbit varies noticeably with different momenta since the dipole field acts only in the y-direction. Fig.~(\ref{dispersion_stage1_before}) shows that dispersion mainly exists in the x-direction, while the y-direction dispersion remains nearly zero. Stage 1 terminates at 104.4 m and connects with a later stage which has a smaller beta function value at the wedge absorber. Stage 4 has the highest $\mathrm{B_z}$ field (7.2 T for the on-axis field) and smallest transverse beta function (23 cm at the wedge absorber) before the bunch merging in order to decrease the transverse and longitudinal emittance of the muon beam to the required values for the bunch merging system (normalized transverse emittance \textasciitilde=1.3 mm and normalized longtudinal emittance \textasciitilde=1.7 mm) \cite{bunch_merging}.  It is worth mentioning that we double the RF frequency from 352 MHz to 704 MHz for stages 3 and 4. This allows for an increase in the RF accelerating gradient and a decrease in the longitudinal beta function, as indicated by Eq. (\ref{longitudinal_beta}). This adjustment contributes to reducing the longitudinal emittance. 

The final emittance, transmission and reference momentum of each stage are listed in Table \ref{emittance_results_before}. The reference momentum refers to the longitudinal momentum of the reference particle, which is used for the timing of the RF cells. The evolution of the transverse and longitudinal emittance is shown in Fig. (\ref{emittance_before}). As depicted in Fig. (\ref{emittance_before}), a significant spike is evident at the junction of stage 3 and stage 4. This spike is due to the longitudinal mismatching as we start to use the 704 MHz RF in the stage 4. In summary, the pre-merging cooling section consists of 4 stages with a total length of 362.8 m. It effectively reduces the transverse and longitudinal emittance from 16.96 mm and 45.53 mm to 1.239 mm and 1.741 mm, respectively, with an overall transmission rate of 49.6\% including the muon decays. The particle distribution in phase spaces at the beginning and end of the cooling section is illustrated in Fig. (\ref{px_x_before}), (\ref{py_y_before}) and (\ref{pz_t_before}). As shown in Fig.~(\ref{px_x_before}), the centers of the initial and final beams are noticeably different. This difference arises because the beam center approximately follows the closed orbit, which reduces as the solenoid focusing increases in later stages. It is encouraging to observe from Fig. (\ref{merit_factor_before}) that the merit factor increases at the end of each stage indicating each cooling stage is well-designed. Similar to the emittance evolution Fig. (\ref{emittance_before}), the merit factor in Fig. (\ref{merit_factor_before}) drops significantly at the start of the stage 3 mostly because of the longitudinal mismatching resulting from the sudden jump in RF frequency from 352 MHz to 704 MHz.
\begin{figure*}[htbp]
  \centering
  \begin{subfigure}[b]{6.45cm}
    \includegraphics[width=6.45cm]{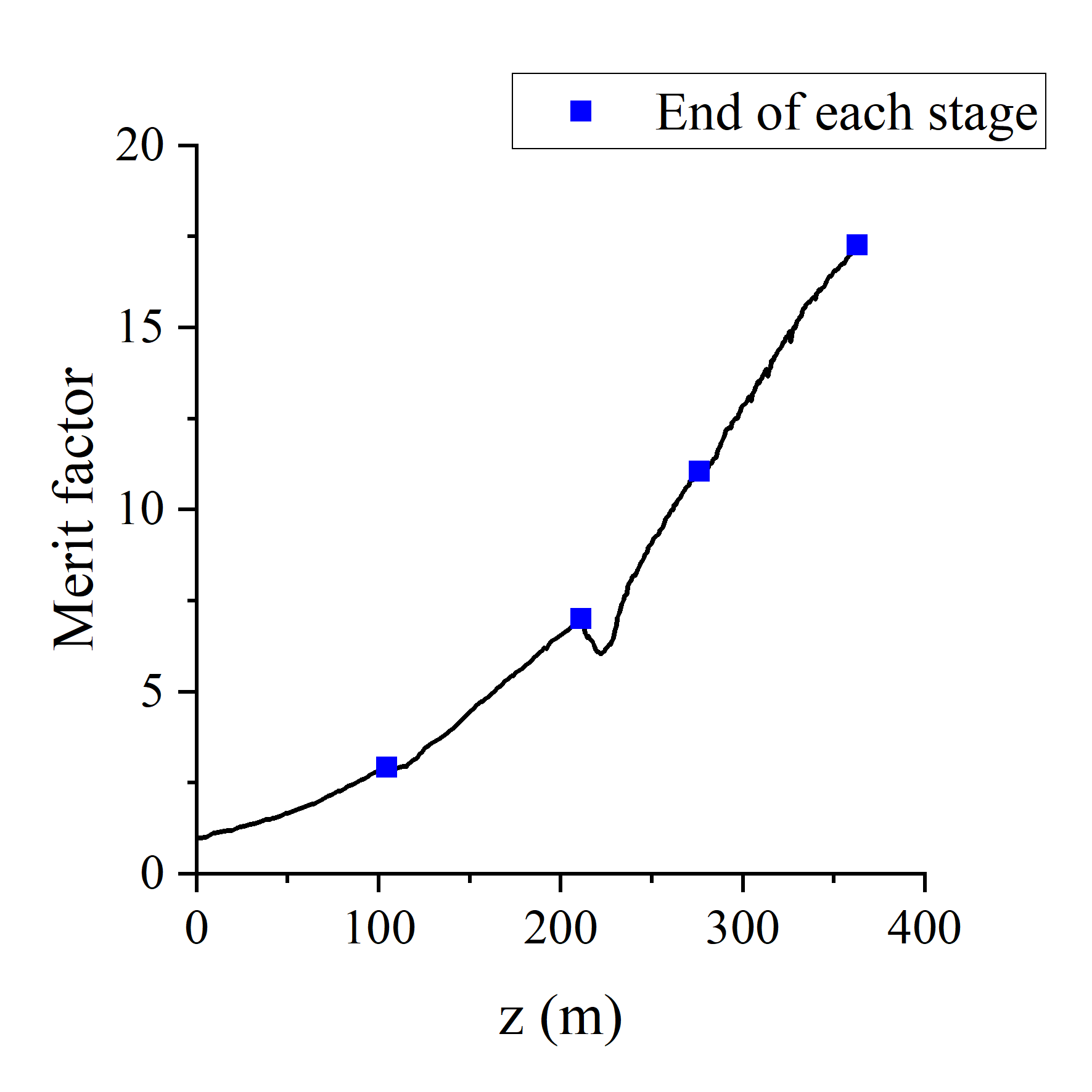}
    \caption{}
    \label{merit_factor_before}
  \end{subfigure}
  \begin{subfigure}[b]{6.45cm}
    \includegraphics[width=6.45cm]{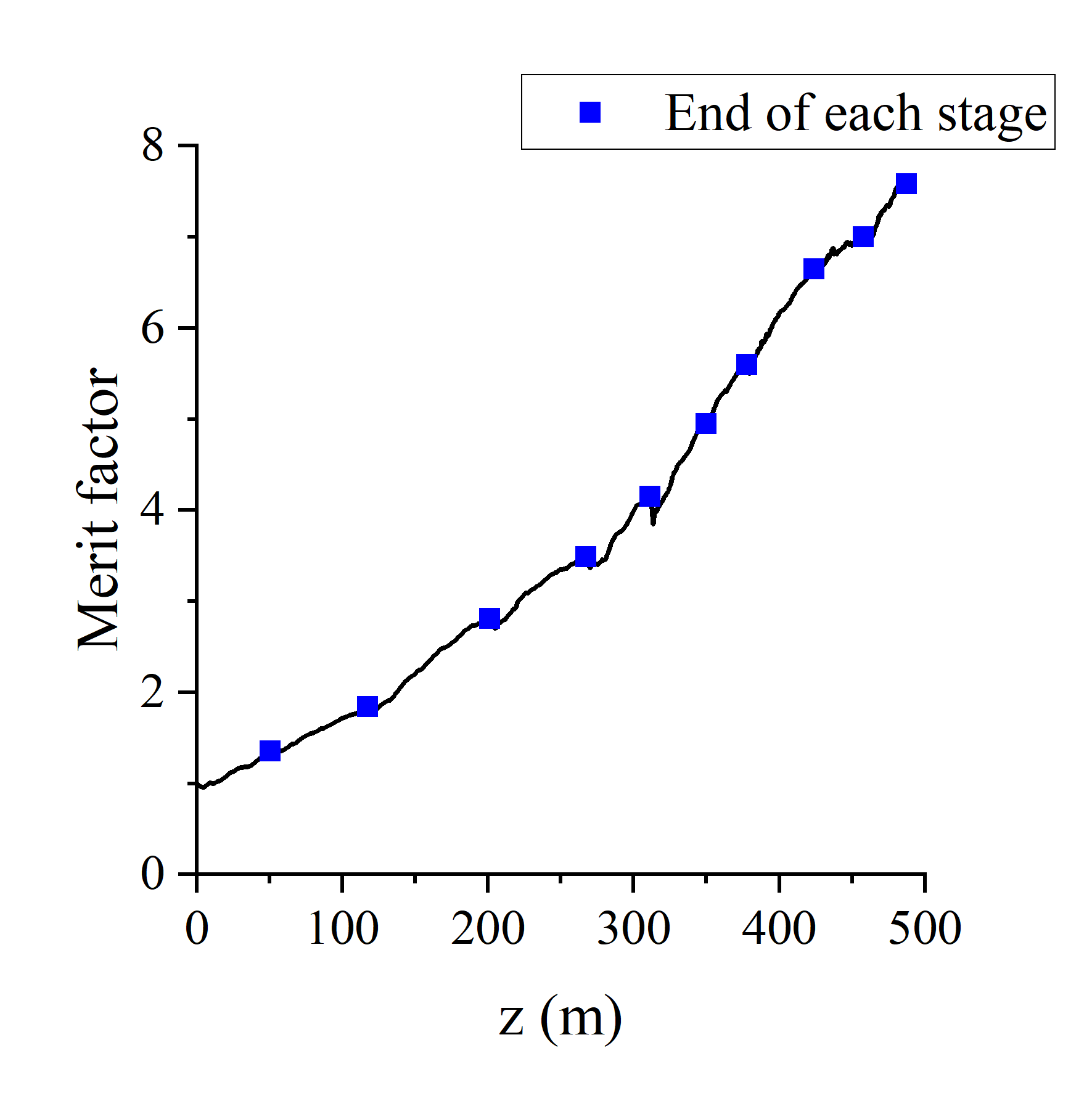}
    \caption{}
    \label{merit_factor_post}
  \end{subfigure}
  \caption{Evolution of the merit factor: (a)  pre-merging section; (b) post-merging section}
  \label{fig_merit_factor}
\end{figure*}
\begin{figure*}[htbp]
  \centering
  \begin{subfigure}[b]{6.45cm}
    \includegraphics[width=6.45cm]{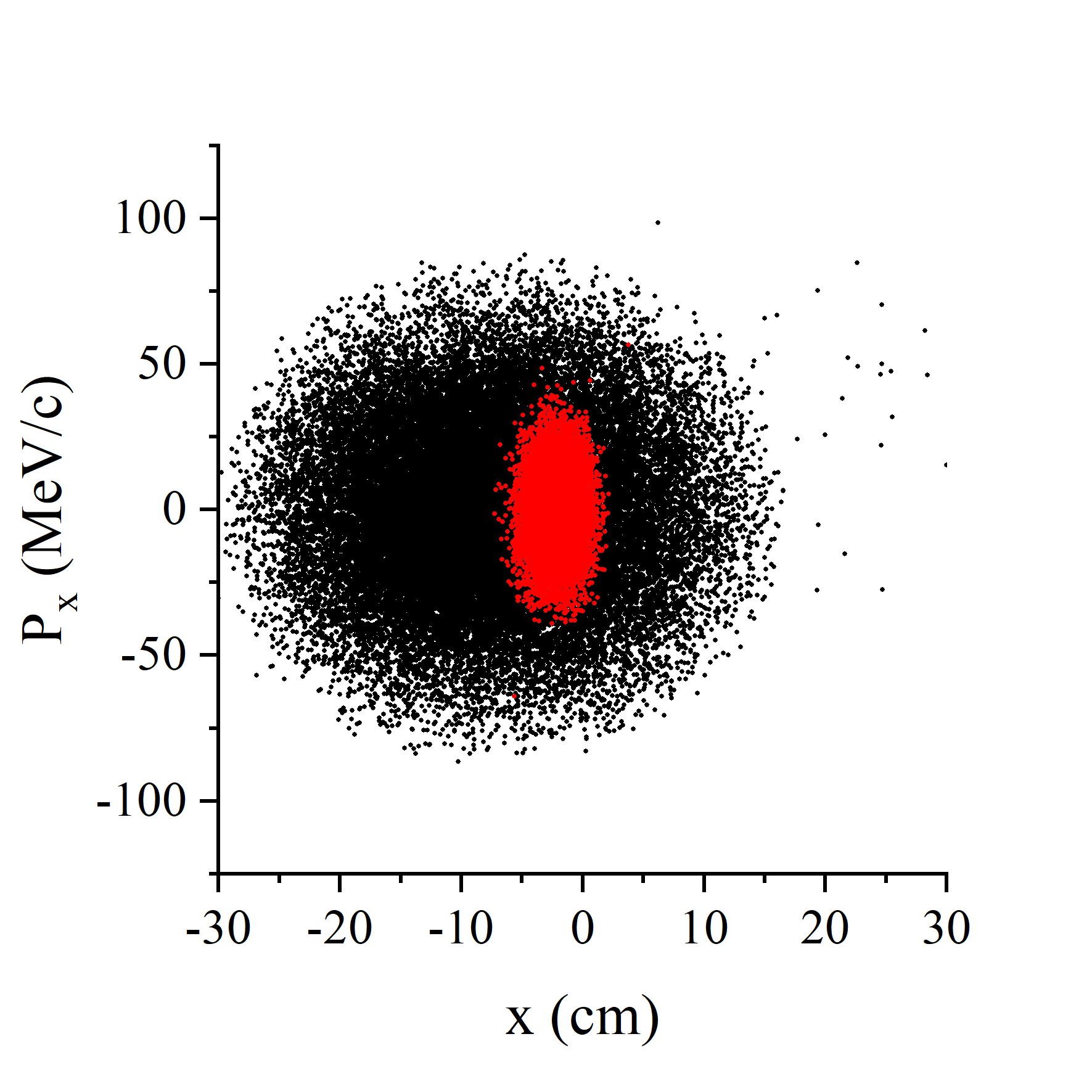}
    \caption{}
    \label{px_x_before}
  \end{subfigure}
  \begin{subfigure}[b]{6.45cm}
    \includegraphics[width=6.45cm]{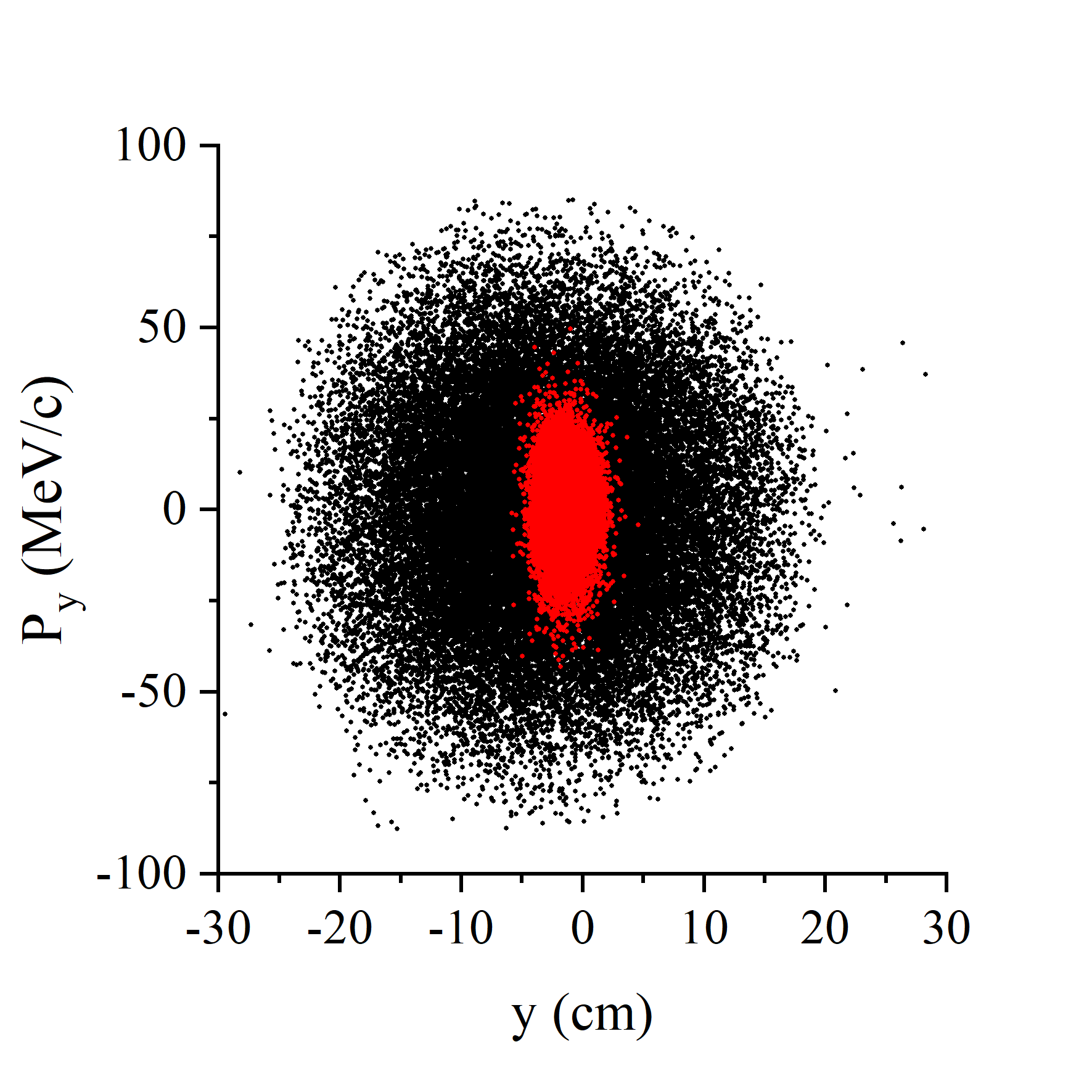}
    \caption{}
    \label{py_y_before}
  \end{subfigure}
  \begin{subfigure}[b]{6.45cm}
    \includegraphics[width=6.45cm]{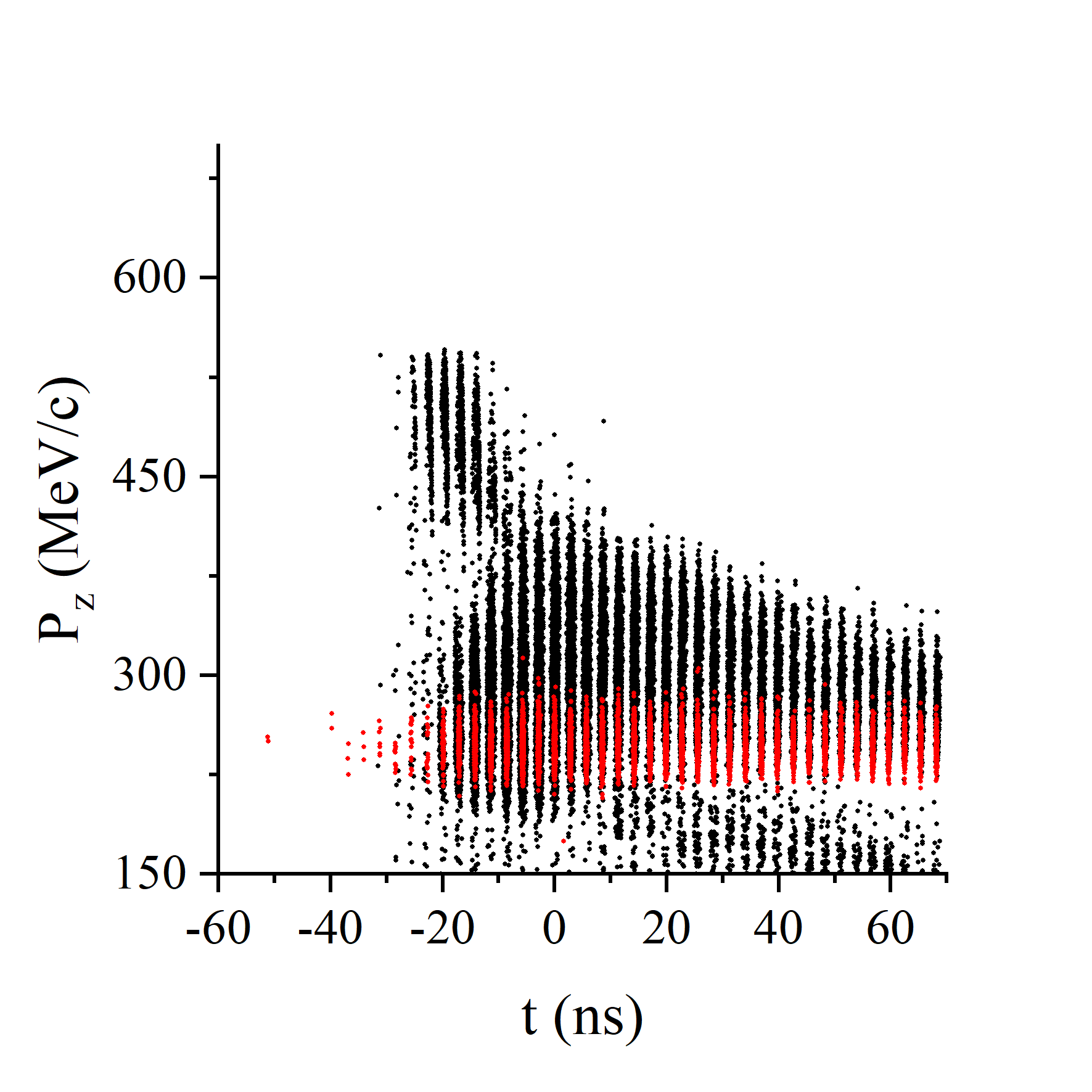}
    \caption{}
    \label{pz_t_before}
  \end{subfigure}
  \begin{subfigure}[b]{6.45cm}
    \includegraphics[width=6.45cm]{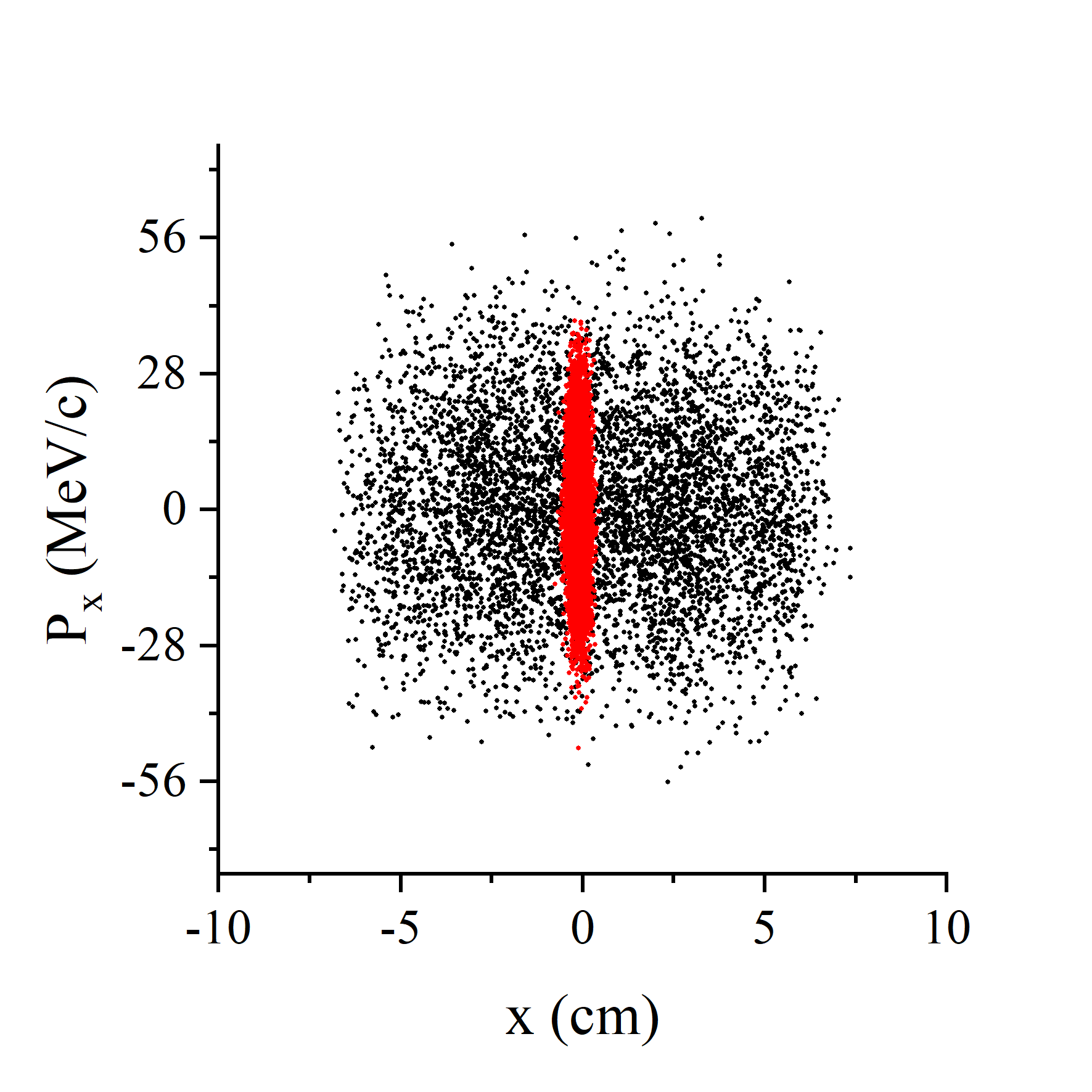}
    \caption{}
    \label{px_x_post}
  \end{subfigure}
  \begin{subfigure}[b]{6.45cm}
    \includegraphics[width=6.45cm]{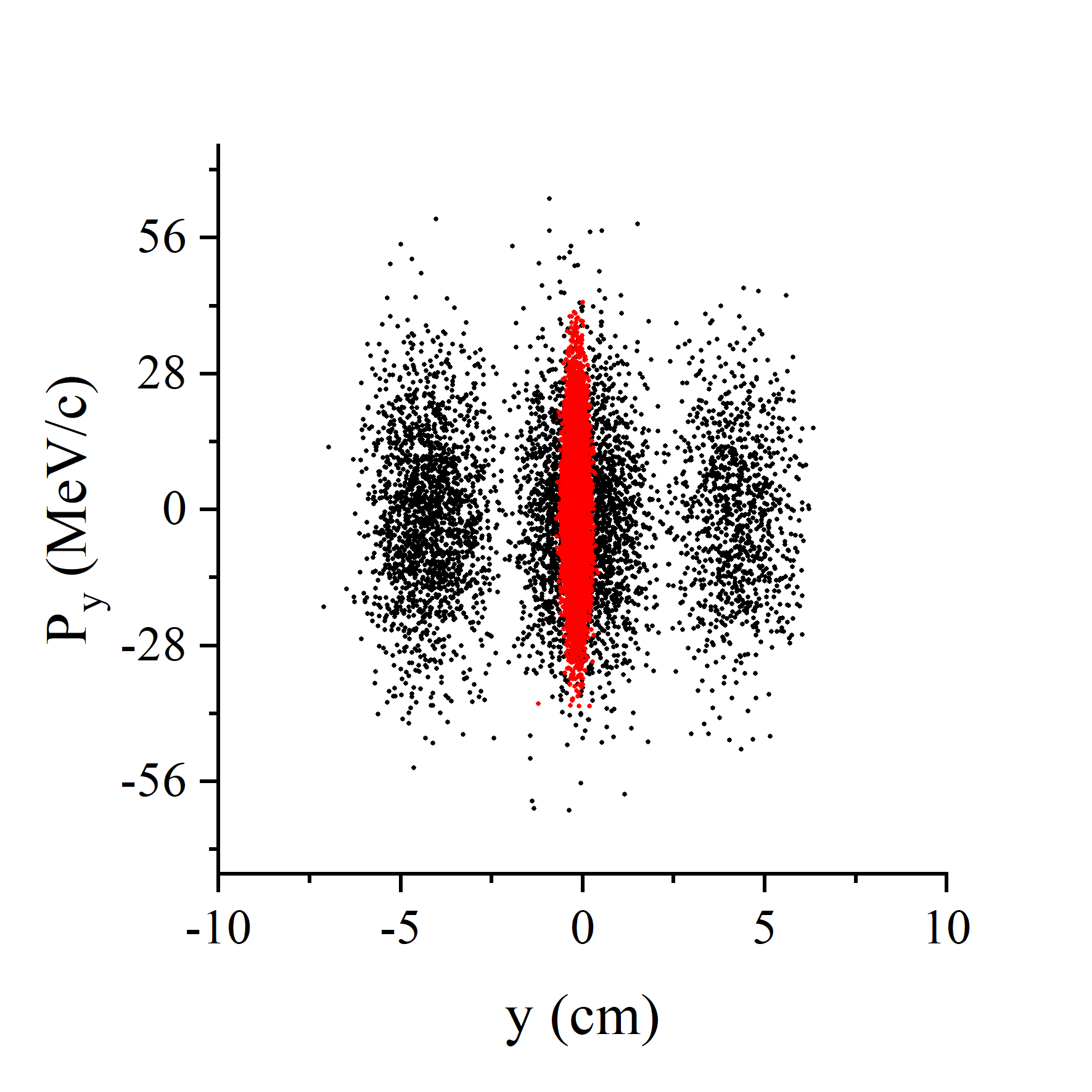}
    \caption{}
    \label{py_y_post}
  \end{subfigure}
  \begin{subfigure}[b]{6.45cm}
    \includegraphics[width=6.45cm]{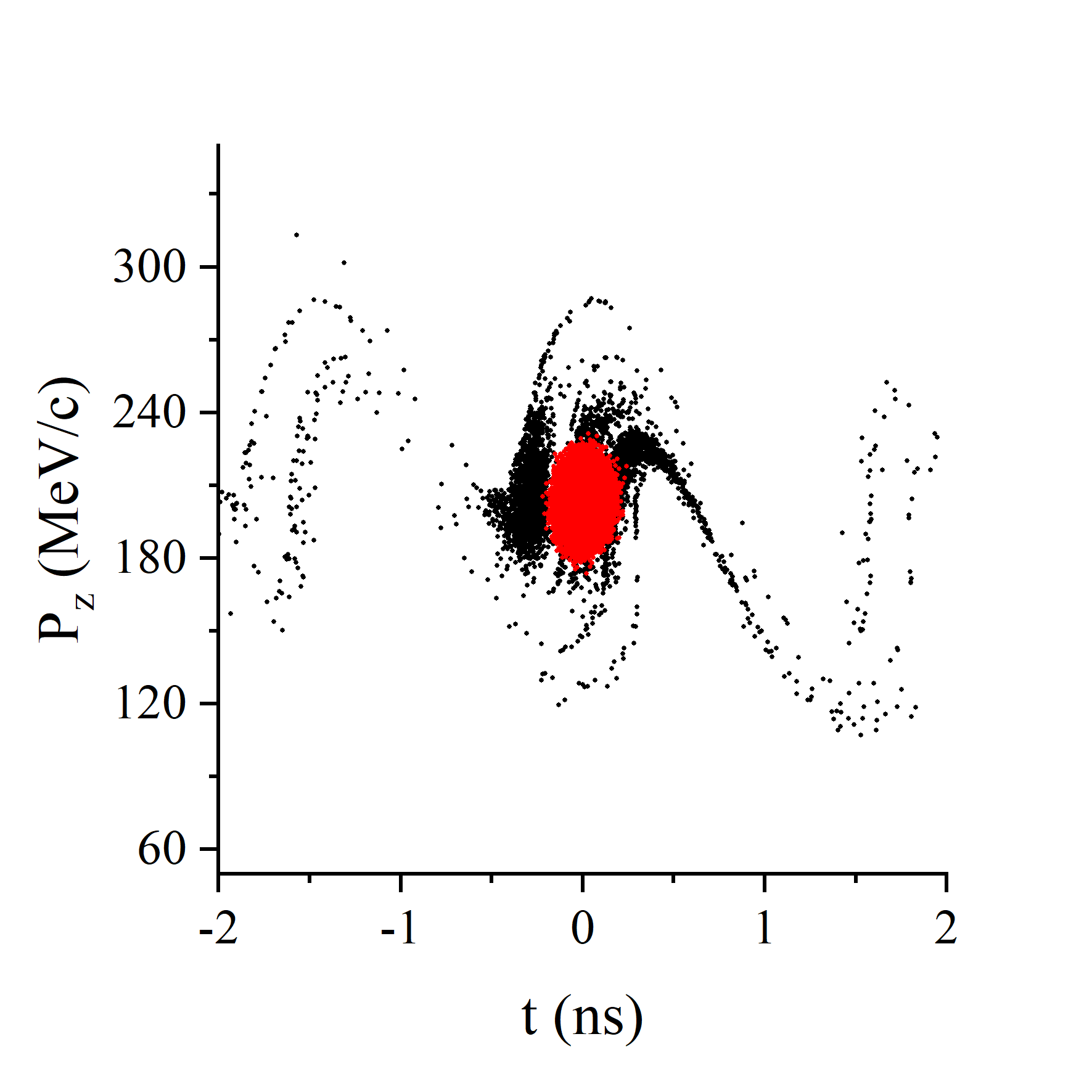}
    \caption{}
    \label{pz_t_post}
  \end{subfigure}
  \caption{Particles distribution in the phase space: (a), (b), (c), pre-merging section; (d), (e), (f), post-merging section (black and red dots denote the distribution at the start and the end of the cooling section respectively)}
  \label{particles_distribution}
\end{figure*}
\subsection{Design of the post-merging cooling section}
After the bunch merging system, both the transverse and longitudinal emittance of the muon beam increase by a factor of \textasciitilde4 \cite{bunch_merging}. We choose to maintain the phase advance of the cooling cell in the post-merging section between $\mathrm{\pi}$ and 2$\mathrm{\pi}$ to achieve a smaller transverse beta function. Despite the fact that this choice results in a narrower longitudinal momentum acceptance, the significantly smaller initial longitudinal emittance in the post-merging section, compared to the pre-merging section, allows for such an approach. As the period of $B_z^2$ for the cooling cells of all stages in this section is the length of the cooling cell, the layout shown in Fig.~(\ref{cell_layout_b}) is adopted for all cooling cells in this section. It is also important to note that a single pair of solenoid coils with opposite polarity is used in each cooling cell from stages 1 to 3. Two pairs are used in stages 4 and 5, and three pairs are employed from stages 6 to 10. The use of multiple pairs of coils helps reduce the current density in the coils. Since the output beam of the post-merging rectilinear cooling section serves as the input beam for the final cooling section through deceleration, our objective is to minimize the emittance of the output beam of the post-merging rectilinear cooling section while mitigating beam loss. In other words, our aim is to ensure that the merit factor defined in Eq.(\ref{merit_factor}) increases along the channel.

A 10-stage rectilinear cooling channel is used in the post-merging section and its main parameters are summarized in Table \ref{cell_parameters}. Given that the initial transverse emittance of this section is approximately one third of that in the pre-merging section, a smaller transverse beta function value of 35 cm is chosen for stage 1. The on-axis field profile of stage 1 is illustrated in Fig. (\ref{field_b}). Figure (\ref{beta_z_stage1_post}) depicts the evolution of the transverse beta function along the cooling cell in stage 1. In comparison with the pre-merging A-stage 1, the curve shape remains similar but has a smaller value at the start and end due to higher $B_z$. Fig. (\ref{beta_pz_stage1_post}) illustrates the transverse beta function and phase advance versus momentum in stage 1. The phase advance of the 150 MeV/c and 238 MeV/c beams approaches 2$\pi$ and $\pi$, respectively, indicating a momentum acceptance range from 150 MeV/c to 238 MeV/c. A shorter cooling cell is used in later stages to achieve tighter focusing and reduce the transverse beta function. However, tighter focusing means the cell has a poorer longitudinal momentum acceptance if the increase in magnetic field strength does not scale with the reduction of the cooling cell length. From Fig.~(\ref{momentum_acceptance_post}), the momentum acceptance generally reduces throughout the stages. Therefore, it is important to gradually decrease the cell length and transverse beta function in each stage to match the momentum spread of the muon beam with the momentum acceptance. The final cooling system requires the input transverse emittance to be less than 0.3 mm which is achieved at the end of the stage 8. However, we find it is still possible to further reduce the emittance by adding two more stages with moderate particle loss. The evolution of emittance and merit factor are shown in Fig.~(\ref{emittance_post}) and Fig.~(\ref{merit_factor_post}), respectively. The particles distribution in the phase space of the post-merging section is displayed in Fig. (\ref{px_x_post}), (\ref{py_y_post}) and (\ref{pz_t_post}). Here, we provide a brief explanation of the initial beam distribution. The bunch merging process occurs in two stages: longitudinal and transverse merging. Initially, three bunches are merged longitudinally into one so that three distinct sets of particles can be seen in the longitudinal phase space shown in Fig. (\ref{pz_t_post}). Following this, seven bunches are merged transversely, with the particle distribution in the x-y plane illustrated in \cite{bunch_merging}, which yields the substructure seen in Figs. (\ref{px_x_post}) and (\ref{py_y_post}). 

We also calculate the theoretical emittance for the end of each stage in the post-merging section from Eqs. (\ref{emittance_evolution}) and (\ref{cooling_length}) and the results are shown in Table \ref{theory_results}. The initial emittance in Eq. (\ref{emittance_evolution}) is equal to the simulated emittance at the end of  each stage. Compared with the simulation results shown in Table \ref{emittance_results_post}, the largest discrepancy between theory and simulation is 23.9\%. This discrepancy between the simulation and theoretical predictions is expected, given that the theory is entirely linear and treats transverse and longitudinal beam motion separately.  In most cases, the theory yields higher output emittance at the end of each stage. This discrepancy arises primarily because the particles with higher amplitudes are lost on the beam pipe, which the theory does not account for.
\begin{table*}
\caption{\label{emittance_results_post}Simulation results on the normalized emittance and transmission at the end of each stage after merging}
\begin{ruledtabular}
\begin{tabular}{cccccc}
\textbf{} & $\varepsilon_{\rm{T,sim}}$ (mm) & $\varepsilon_{\rm{L,sim}}$ (mm) & $\varepsilon_{\rm{6D,sim}}$ (mm$^3$) & Transmission & Reference momentum (MeV/c)\\
\hline
Start & 5.129& 9.991& 262.5&\\
B-Stage 1& 2.881& 9.085& 76.& 85.2\%& 202\\
B-Stage 2& 1.991& 6.579& 26.7& 89.4\%& 201\\
B-Stage 3& 1.271& 4.05& 6.734& 87.5\%& 203\\
B-Stage 4& 0.9353& 3.161& 2.828& 89.8\%& 202\\
B-Stage 5& 0.7034& 2.511& 1.322& 89.4\%& 205\\
B-Stage 6& 0.4823& 2.291& 0.5527& 88.4\%& 204\\
B-Stage 7& 0.3855& 2.064& 0.312& 92.8\%& 203\\
B-Stage 8& 0.2649& 1.861& 0.1307& 87.9\%& 200\\
B-Stage 9& 0.19& 1.716& 0.0613& 85.2\%& 197\\
B-Stage 10& 0.1396& 1.558& 0.03009& 87.1\%& 200\\
\end{tabular}
\end{ruledtabular}
\end{table*}

In summary, the 10-stage post-merging cooling section is able to reduce the normalized transverse and longitudinal emittance of the muon beam from 5.129 mm and 9.991 mm to 0.1396 mm and 1.558 mm, respectively. The channel length is 487.26 m with the transmission of 28.5\% including the muon decays. The output transverse emittance of this updated design is half that of the previous design \cite{MAP}. A lower initial transverse emittance is always beneficial for the final cooling. Previous simulation studies on final cooling  have successfully reduced the normalized transverse emittance of the muon beam to 55  $\mu$m \cite{final_cooling}. The current studies initiated by the International Muon Collider Collaboration (IMCC) aspire to surpass this achievement, aiming for a value as low as 25 $\mu$m \cite{Towards_a_Muon_Collider}. We anticipate that the output muon beam of this updated post-merging rectilinear cooling channel design will significantly facilitate the final cooling system to achieve the goal of a normalized transverse emittance of 25 $\mu$m. 
\begin{table}
\caption{\label{theory_results}Theory results on the normalized emittance at the end of each stage in the post-merging section and error between the theory and simulation}
\begin{ruledtabular}
\begin{tabular}{ccccc}
\textbf{} & $\varepsilon_{\rm{T,theo}}$ (mm) & Err$_T$ & $\varepsilon_{\rm{L,theo}}$ (mm) & Err$_L$\\ \hline
B-Stage 1& 3.119& 7.6\%& 8.614& 5.5\%\\
B-Stage 2& 2.616& 23.9\%& 5.338& 23.2\%\\
B-Stage 3& 1.665& 23.7\%& 3.559& 13.8\%\\
B-Stage 4& 1.113& 16\%& 2.967& 6.5\%\\
B-Stage 5& 0.8083& 13\%& 2.568& 2.2\%\\
B-Stage 6& 0.5099& 5.4\%& 2.540& 9.8\%\\
B-Stage 7& 0.4158& 7.3\%& 2.130& 3.1\%\\
B-Stage 8& 0.3442& 23\%& 1.703& 9.3\%\\
B-Stage 9& 0.2477& 23.3\%& 1.659& 3.4\%\\
B-Stage 10& 0.1832& 23.8\%& 1.647& 5.4\%\\
\end{tabular}
\end{ruledtabular}
\end{table}
\section{Tracking simulation using $\pi$-mode RF and error analysis}\label{section_5}
When a particle traverses the RF structures of the cooling channel, the phasing between adjacent cells is required to match the time of flight across each cell. The RF cell length is related to the frequency and particle velocity according to 
\begin{equation}\label{RF_mode_length}
L=\frac{{\beta}c}{2{\pi}f} \Delta \phi_{RF}
\end{equation}
where L is the length of a RF cell, $\Delta \phi_{RF}$ is the relative phase between adjacent RF cells (e.g., $\pi$/2, $\pi$...), ${\beta}c$ is the velocity of the muon beam and $f$ is the RF frequency. In the lattice described in section \ref{section_4}, adjacent cells have a phase difference around $\pi$/2. Each cavity is expected to have an individual power coupler and feed-through to the cavity. 

In order to simplify the engineering, coupled RF cells with a phase difference of $\pi$ can be used. In this case, one power coupler and feed-through is required for an entire structure, with adjacent cells coupled either through the iris or through the cavity walls. In order to achieve a correct phasing of the beam, their length must be doubled compared to that of $\pi$/2-mode RF cells. 

The transit time factor describes the degradation in effective voltage resulting from phase variation of the cavity during passage of the beam across the cavity. The transit time factor is given by
\begin{equation}\label{transit_time_factor}
T=\frac{\sin{\omega_{rf}L/2{\beta}c}}{\omega_{rf}L/2{\beta}c}
\end{equation}
where $\omega_{rf}$ is the angular frequency of the RF, L is the length of the RF and ${\beta}c$ is the speed of the reference particle. If the length of the cavity is extended, the transit time factor is reduced so that higher  RF gradients are required to restore the energy loss in the absorber, as seen from Eq. (\ref{energy_gain}).
\begin{equation}\label{energy_gain}
    \Delta{E_{rf}}=N_{rf}TV_{rf}L\sin{\varphi_s}
\end{equation}
where $\Delta{E_{rf}}$ is the energy gain from the RF, $N_{rf}$ is the number of RF cells, T is the transit time factor, L is the length of  each RF cell and $\varphi_s$ is the RF phase. 
\begin{table*}
\caption{\label{pi_mode_cell_parameters}Main parameters of the cooling cell for $\pi$-mode and normal mode RF}
\begin{ruledtabular}
\begin{tabular}{ccccccccccccccc}
\textbf{ } &\makecell[c]{Cell \\length \\(m)} &  \makecell[c]{Stage \\length \\(m)}&\makecell[c]{Max. \\ on-axis \\ $\mathrm{B_z}$ (T)}  & \makecell[c]{Integrated\\ $\mathrm{B_y}$ (T·m)}& \makecell[c]{Transverse \\beta \\(cm)}& \makecell[c]{Dispersion \\(mm)}& \makecell[c]{On-axis \\wedge \\length \\(cm)}& \makecell[c]{Wedge \\apex \\angle \\(deg)}& \makecell[c]{RF \\frequency \\(MHz)}& \makecell[c]{Number \\of \\RF cells} & \makecell[c]{RF cell\\length \\(cm)} &\makecell[c]{Transit \\ time \\ factor}& \makecell[c]{Max. \\RF \\gradient\\ (MV/m)} & \makecell[c]{RF \\phase \\(deg)}\\ \hline
$\pi$-mode & 0.9&  44.1&8.1&0.105& 10& -17.5& 12& 120& 704& 3& 18.8 &0.63& 26.3& 23.8\\
normal mode& 0.9&  44.1&8.1&0.105& 10& -17.5& 12& 120& 704& 5& 9.5 &0.90& 22.3&23.0\\
\end{tabular}
\end{ruledtabular}
\end{table*}
\begin{table}
\caption{\label{emittance_results_pi}Simulation results on the normalized emittance and transmission of the normal and $\pi$-mode lattice}
\begin{ruledtabular}
\begin{tabular}{ccccc}
\textbf{} & $\varepsilon_{\rm{T,sim}}$ (mm) & $\varepsilon_{\rm{L,sim}}$ (mm) & $\varepsilon_{\rm{6D,sim}}$ (mm$^3$) & Transmission\\
\hline
Start& 0.9314& 3.200& 2.793&\\
$\pi$-mode& 0.7082& 2.569& 1.367&90.0\%\\ 
Normal mode& 0.6996& 2.571& 1.325&89.6\%\\
\end{tabular}
\end{ruledtabular}
\end{table}
\begin{figure*}[htbp]
  \centering
  \begin{subfigure}[b]{6.45cm}
    \includegraphics[width=6.45cm]{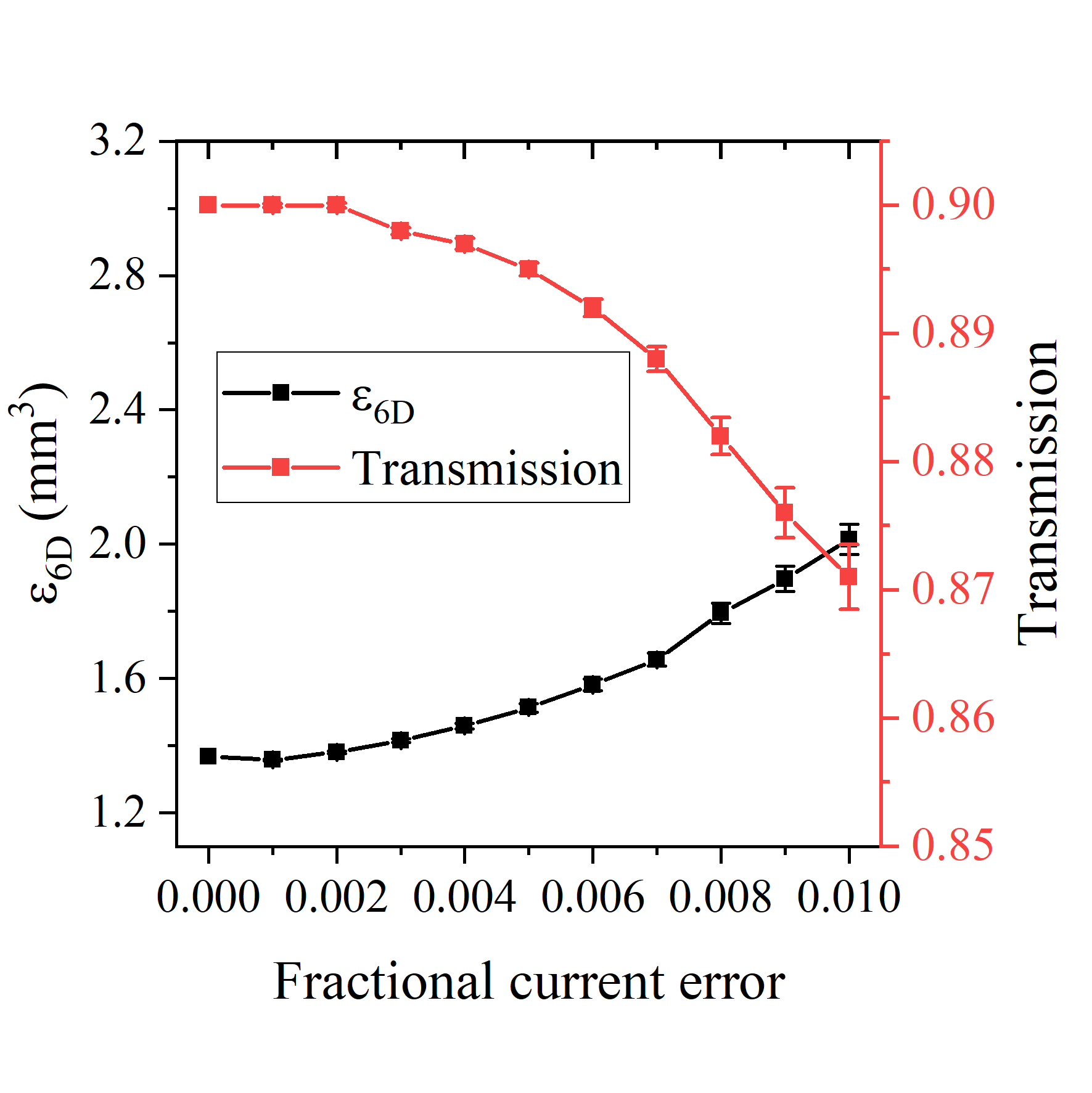}
    \caption{}
    \label{solenoid_current_error}
  \end{subfigure}
  \begin{subfigure}[b]{6.45cm}
    \includegraphics[width=6.45cm]{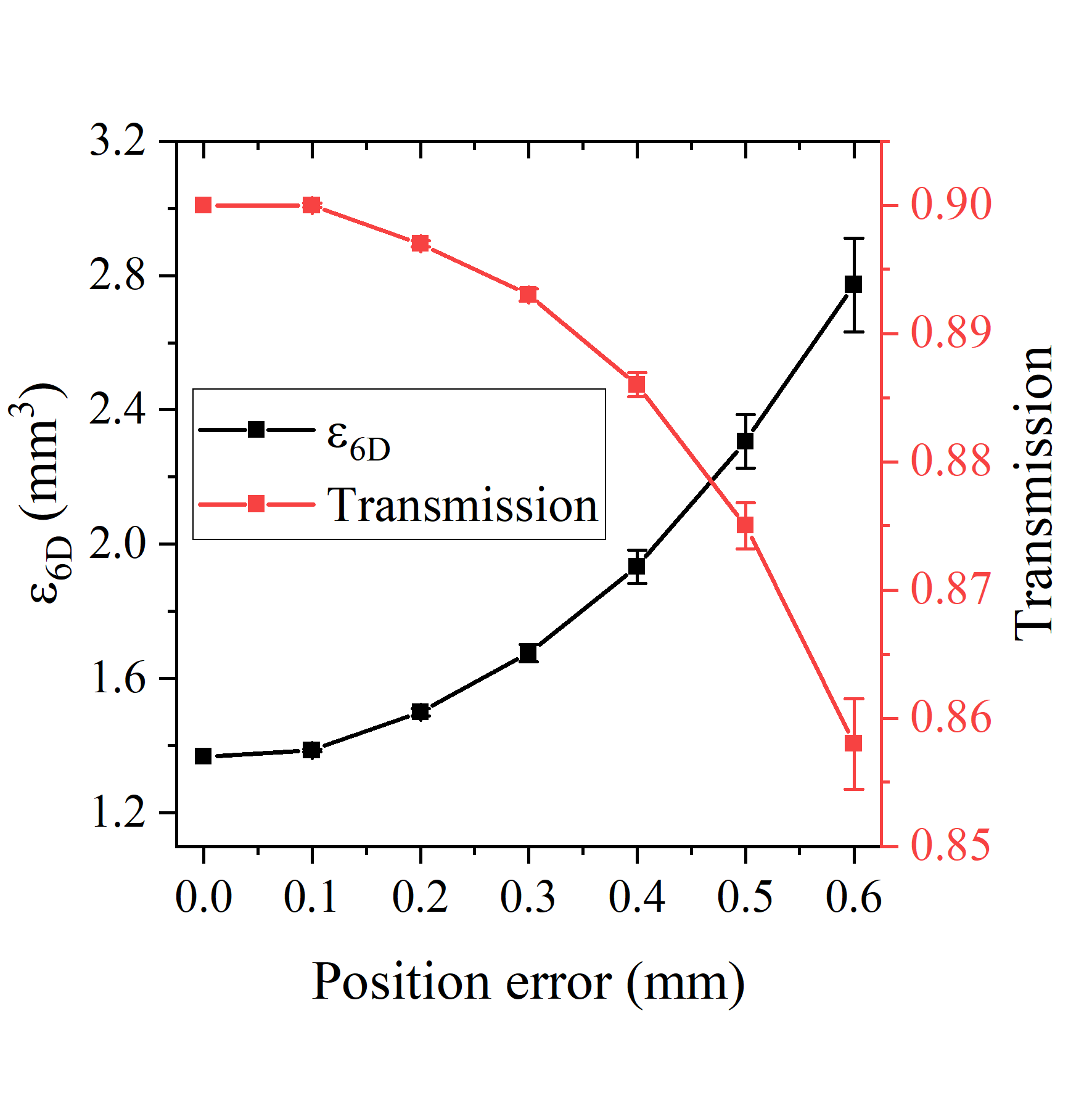}
    \caption{}
    \label{solenoid_position_error}
  \end{subfigure}
  \begin{subfigure}[b]{6.45cm}
    \includegraphics[width=6.45cm]{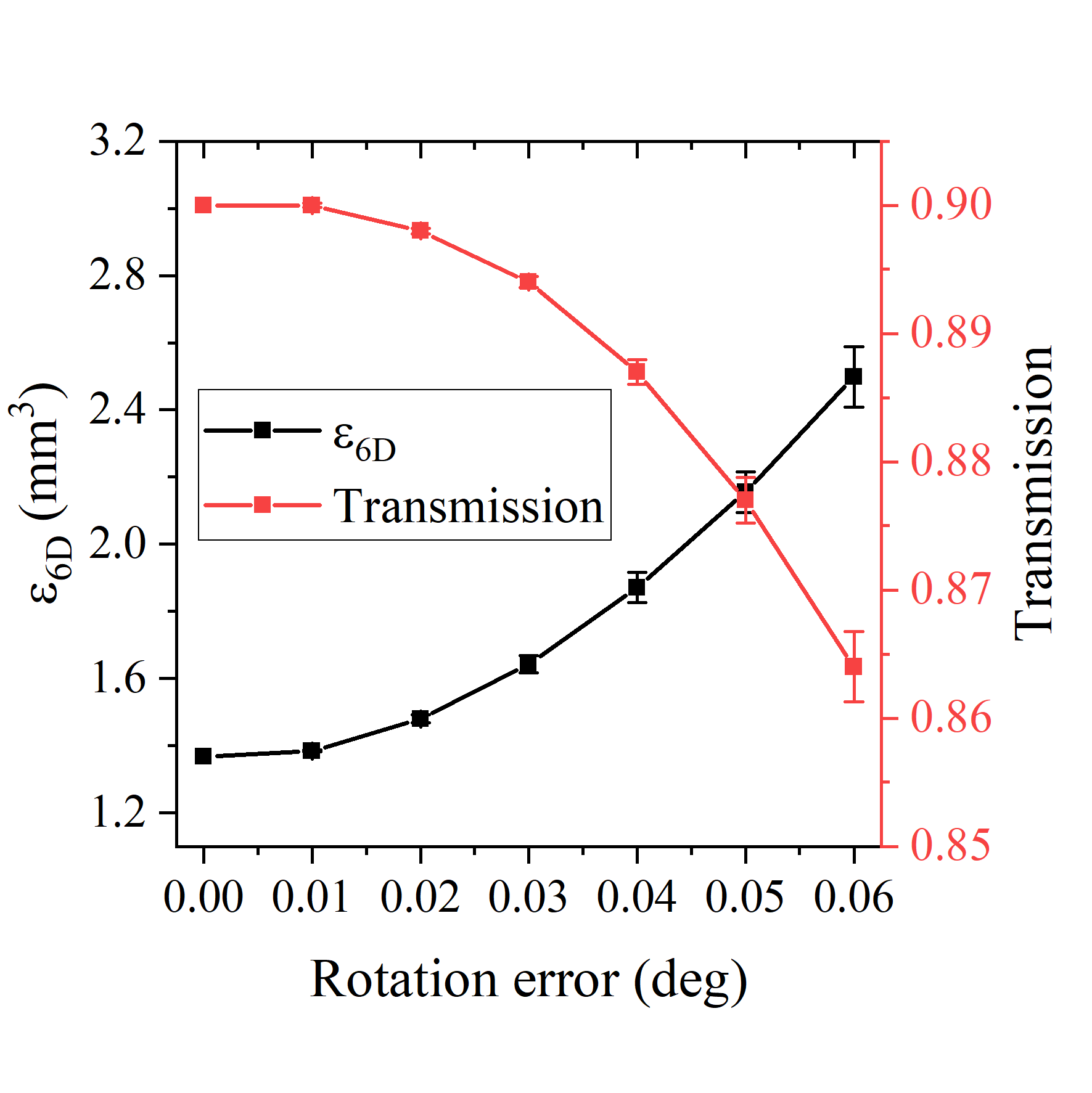}
    \caption{}
    \label{solenoid_rotation_error}
  \end{subfigure}
  \begin{subfigure}[b]{6.45cm}
    \includegraphics[width=6.45cm]{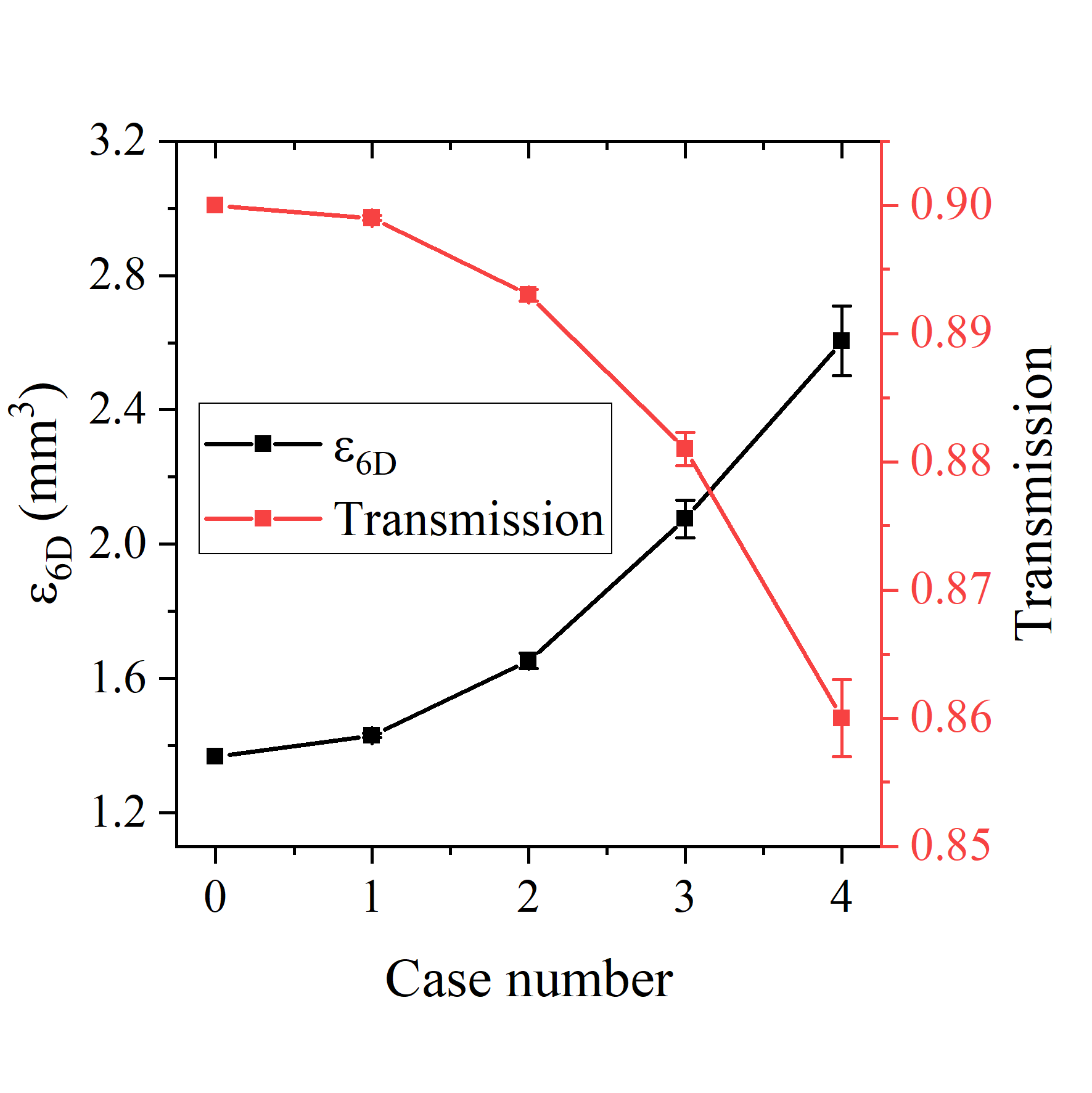}
    \caption{}
    \label{solenoid_current_position_rotation_error}
  \end{subfigure}
  \caption{Variation of normalized 6D emittance and transmission due to different types of solenoid coil errors: (a) fractional current errors; (b) position errors; (c) rotation errors; (d) combined current, position, and rotation errors}
  \label{solenoid_error}
\end{figure*}
\begin{figure*}[htbp]
  \centering
  \begin{subfigure}[b]{6.45cm}
    \includegraphics[width=6.45cm]{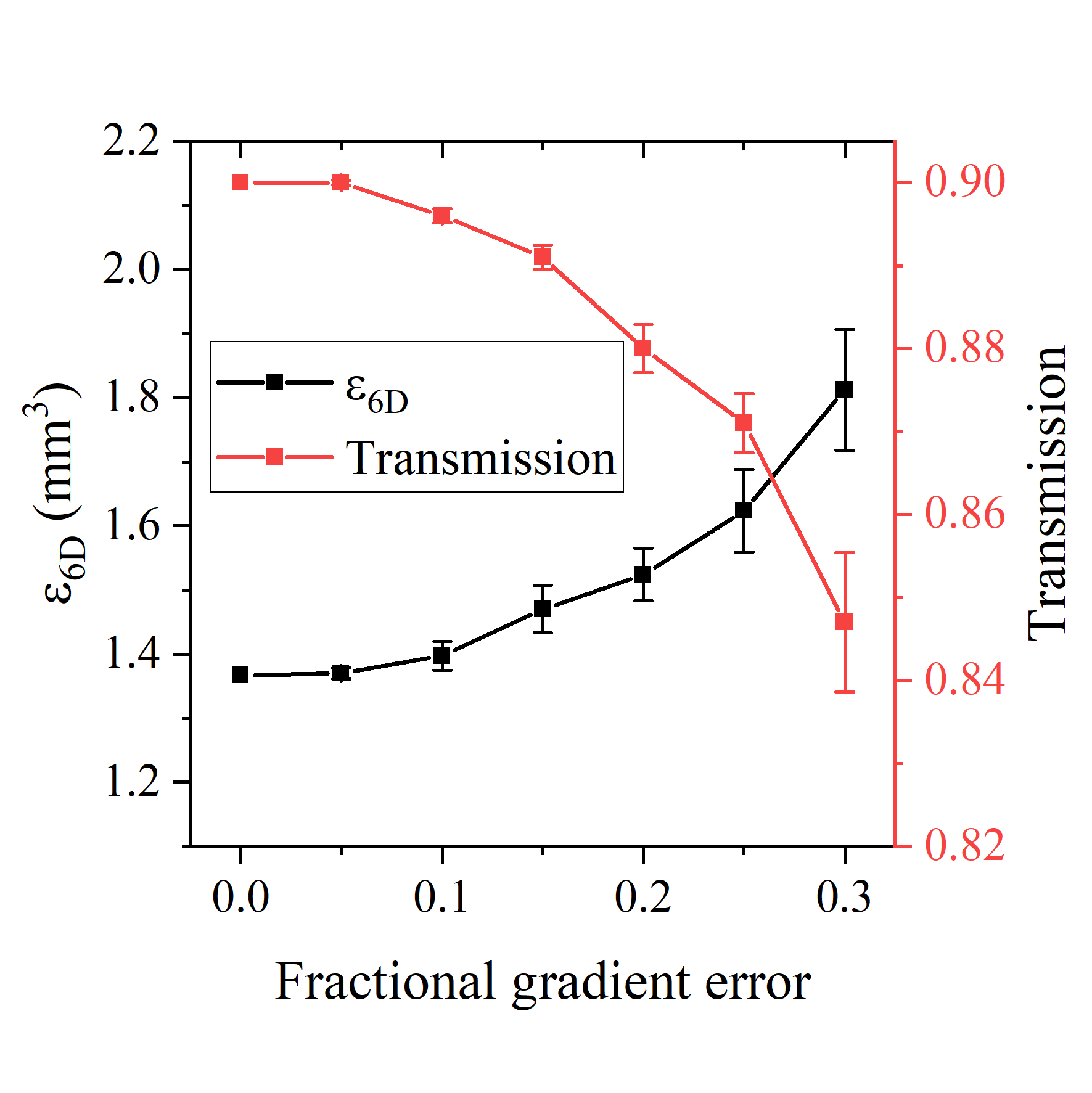}
    \caption{}
    \label{rf_gradient_error}
  \end{subfigure}
  \begin{subfigure}[b]{6.45cm}
    \includegraphics[width=6.45cm]{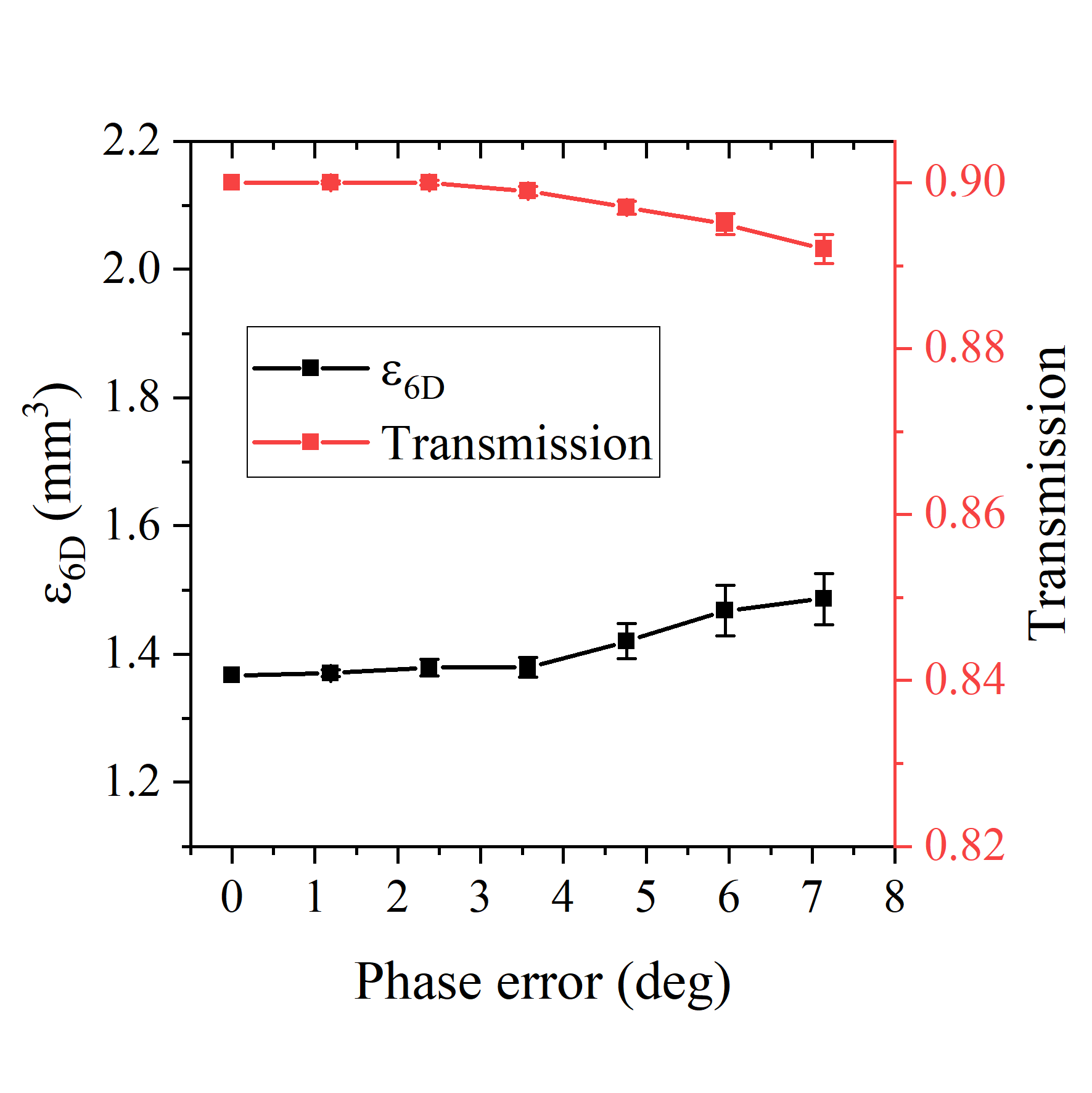}
    \caption{}
    \label{rf_phase_error}
  \end{subfigure}
  \begin{subfigure}[b]{6.45cm}
    \includegraphics[width=6.45cm]{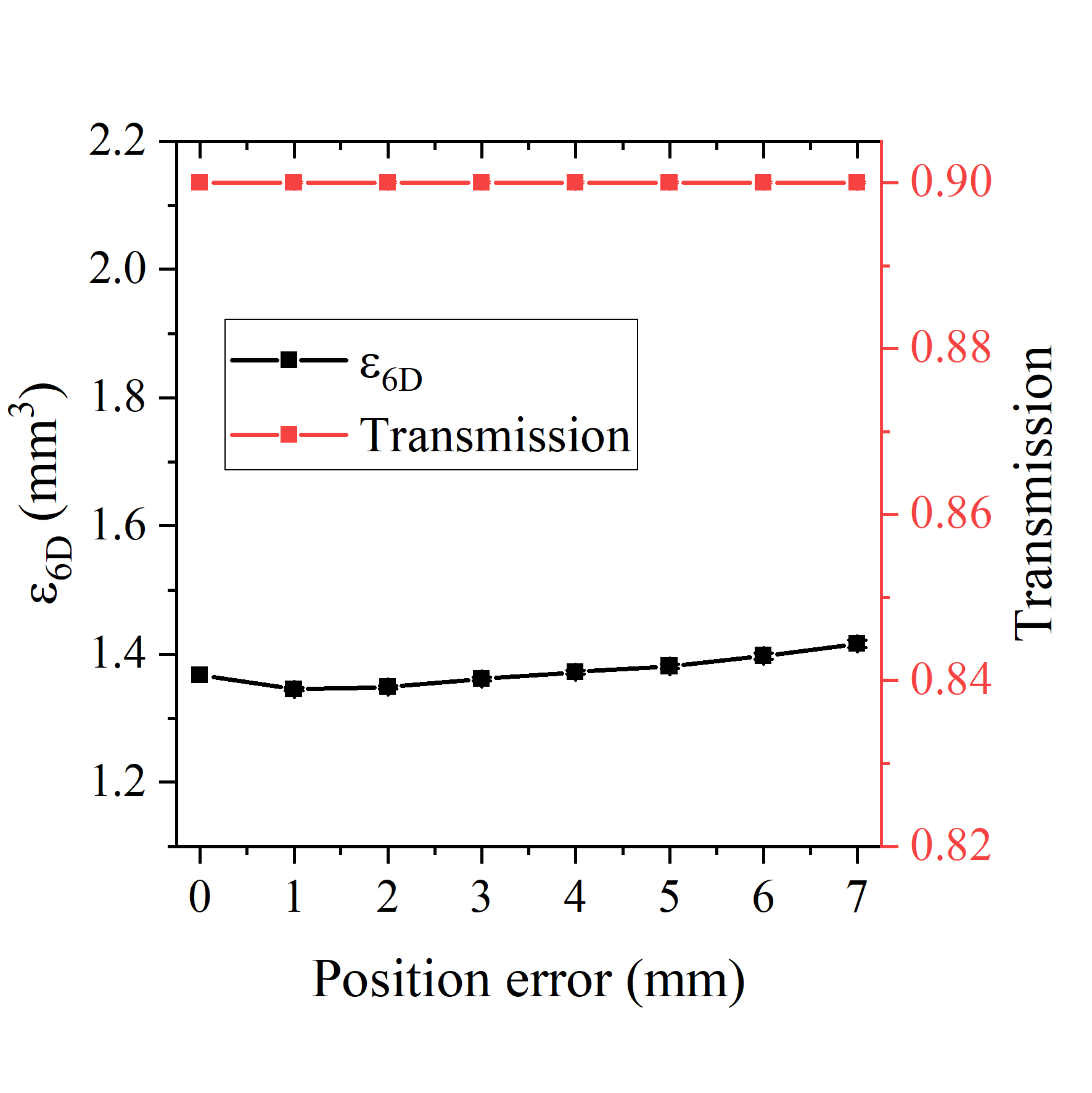}
    \caption{}
    \label{rf_position_error}
  \end{subfigure}
  \begin{subfigure}[b]{6.45cm}
    \includegraphics[width=6.45cm]{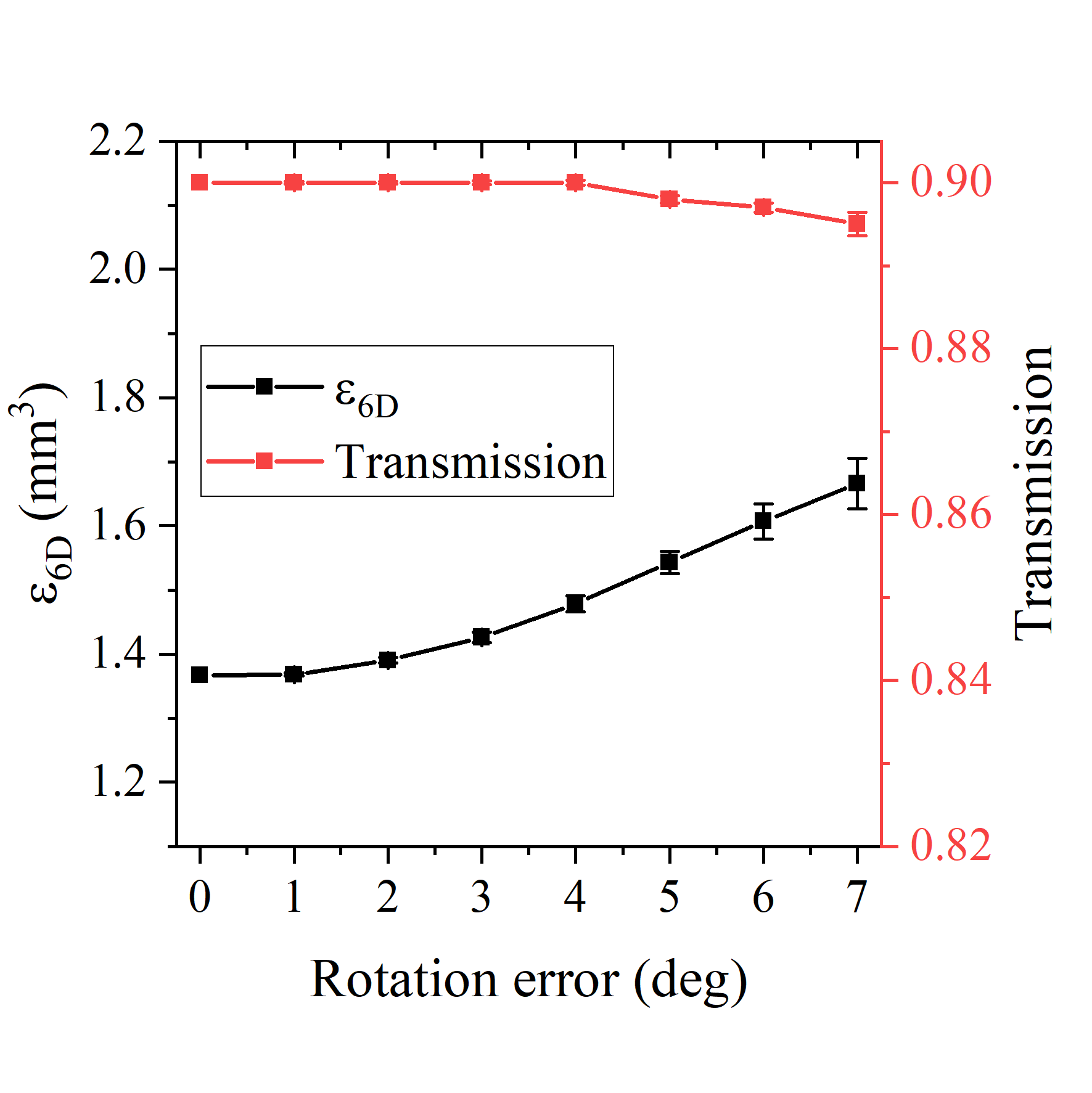}
    \caption{}
    \label{rf_rotation_error}
  \end{subfigure}
  \caption{Variation of normalized 6D emittance and transmission due to different types of RF errors: (a) fractional gradient errors; (b) phase errors; (c) position errors; (d) rotation errors}
  \label{rf_error}
\end{figure*}
\begin{figure}[htbp]
  \centering
  \includegraphics[width=\columnwidth]{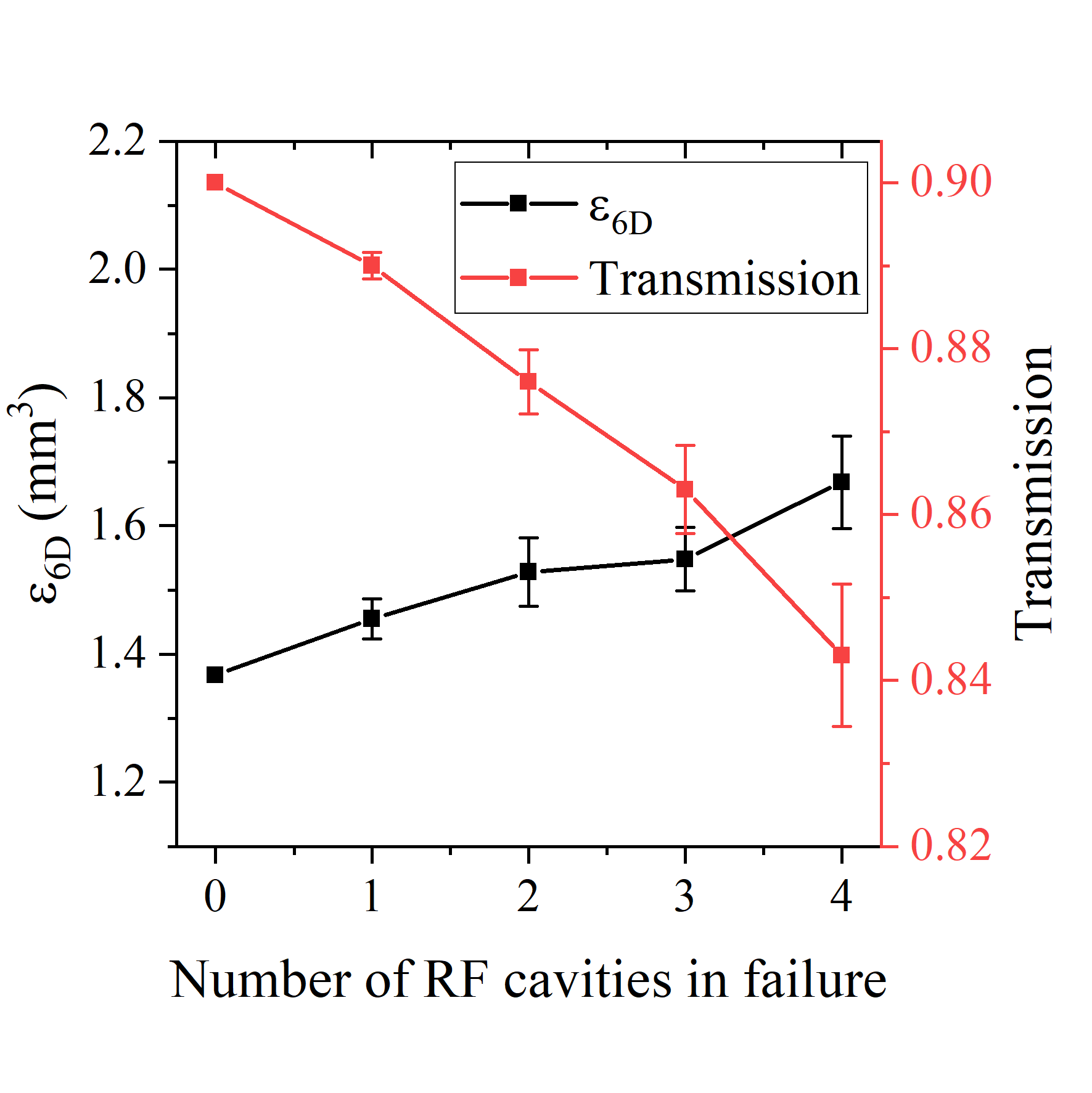}
  \caption{Variation of normalized 6D emittance and transmission due to RF failure}
  \label{rf_failure}
\end{figure}

Due to the simplicity of the power couplers, the International Muon Collider Collaboration will use $\pi$-mode RF for a 6D muon cooling demonstrator. As the initial emittance of stage 5 in the post-merging cooling section is similar to that in \cite{demonstrator}, we choose the cooling cell in stage 5 as a baseline to check the impact of $\pi$-mode RF on cooling performance. Since the length of one $\pi$-mode RF cell is 18.8 cm, nearly double that of the normal mode RF cell discussed in Section \ref{section_4}, we have increased the length of the cooling cell from 80 cm to 90 cm. The lattice parameters and tracking results of the emittance and transmission for the two cases of $\pi$-mode and normal mode RF are summarized in Table \ref{pi_mode_cell_parameters} and Table \ref{emittance_results_pi}, respectively. In order to only investigate if $\pi$-mode RF influences the cooling performance, we maintain identical magnetic field and wedge absorber settings in both cases, with the only differences being in the RF length, gradient, and phase. The tracking results in Table \ref{emittance_results_pi} indicate that there is no obvious difference in cooling performance between the normal and $\pi$-mode cases. From Table \ref{pi_mode_cell_parameters}, it is evident $\pi$-mode RF has higher peak gradient and phase compared with the normal mode, which is due to a lower transit time factor. 

To understand the robustness of the cooling lattice, error analysis studies are conducted. Since the beam emittance and lattice parameters differ in each cooling stage, detailed results on error analysis would vary. For convenience, this analysis is performed on the $\pi$-mode lattice in this paper. Error analysis for other cooling stages will be conducted in future work. 

Two sources of errors are considered: those originating from the solenoid coils and the RF cells. For the solenoid coils, errors are classified into three types: current, position, and rotation. For the RF cells, errors are classified into four types: gradient, phase, position and rotation. The steps of a simulation for a specific error are as follows: (a) Generate random numbers (errors) from a truncated Gaussian distribution truncated at 3 standard deviations from the mean.; (b) Apply these errors to the solenoid coils or RF cells in the simulation. (c) Repeat step (a) and (b) for 100 iterations. (d) Average the output emittance and transmission from these 100 simulation results. For clarity, values of different types of errors in the following paragraphs and Figs. (\ref{solenoid_error}), (\ref{rf_error}) and (\ref{rf_failure}) denote the RMS values used to generate these random errors. The error bars in Figs. (\ref{solenoid_error}), (\ref{rf_error}) and (\ref{rf_failure}) represent the 95$\%$ confidence interval of the estimated mean values.

Fig.~(\ref{solenoid_error}) illustrates the variation in normalized 6D emittance and transmission due to different types of errors in the solenoid coils. Fig.~(\ref{solenoid_current_error}) addresses current errors, with the horizontal axis representing percentage changes in the solenoid coil current relative to the nominal values. Fig.~(\ref{solenoid_position_error}) focuses on position errors, with errors added to both transverse (x and y) and longitudinal (z) positions of the coils in the simulations. Fig.~(\ref{solenoid_rotation_error}) addresses rotation errors, where the coils are randomly rotated around the x, y, and z axes. Fig.~(\ref{solenoid_current_position_rotation_error}) depicts the impact of combined current, position, and rotation errors in the solenoid coils, with the horizontal axis representing five cases corresponding to different error magnitudes: (0\%, 0 mm, 0°), (0.1\%, 0.1 mm, 0.01°), (0.2\%, 0.2 mm, 0.02°), (0.3\%, 0.3 mm, 0.03°), and (0.4\%, 0.4 mm, 0.04°). It can be estimated from Figs.~(\ref{solenoid_current_error}), (\ref{solenoid_position_error}) and (\ref{solenoid_rotation_error}) that, when only one type of error is applied, the thresholds for current, position, and rotation errors to begin noticeably degrading the cooling performance (with a transmission reduction >=1\%) are approximately 0.6\%, 0.3 mm, and 0.03°, respectively. For the combined errors, it can be seen from Fig.~(\ref{solenoid_current_position_rotation_error}) that the threshold is approximately (0.2\%, 0.2 mm, 0.02°). It should be noted that position and rotation errors deserve the most attention, as it is possible for these errors to reach the threshold values in reality. This suggests that a correction scheme may be necessary. However, controlling current errors to under 0.2\% is entirely feasible.

Fig.~(\ref{rf_error}) illustrates the variation in normalized 6D emittance and transmission due to different types of errors in the RF cavities. Compared to the errors in the solenoid coils, the errors in the RF cavities are made deliberately much larger in the simulations in order to show the degradation in cooling performance caused by RF errors. Fig.~(\ref{rf_gradient_error}) shows gradient errors, with the horizontal axis representing percentage changes in the RF gradient relative to the nominal values. When gradient error reaches a very large value, around 10$\%$, it begins to impact the cooling performance. Fig.~(\ref{rf_phase_error}) shows phase errors. The emittance is less sensitive to phase errors compared to gradient errors, with the transmission reduced by only about 1$\%$ when the phase error reaches about 7°. Fig.~(\ref{rf_position_error}) depicts position errors, with errors added to both transverse (x and y) and longitudinal (z) positions of the RF cavities. Even when the position error reaches an extreme value of 7 mm, the transmission remains unaffected, and the 6D emittance increases by only about 4$\%$. Fig.~(\ref{rf_rotation_error}) shows rotation errors, where the RF cavities are randomly rotated around the x, y, and z axes. Similar to the case of position errors, the transmission remains largely unaffected by rotation errors. In practice, RF errors are typically not as large as those depicted in Fig.~(\ref{rf_error}). The purpose of scanning to very large errors is to demonstrate the excellent robustness of the cooling performance against RF errors. The influence on cooling performance caused by RF failure is also simulated, as shown in Fig.~(\ref{rf_failure}). Here, "one RF cavity" refers to all RF cells within a single $\pi$-mode RF cavity. Since $\pi$-mode RF cells are coupled, a failure in one RF cell causes the entire RF cavity to fail. Simulations show that in the case of one RF cavity failure, the 6D emittance increases by about 6\% and the transmission decreases by 1\%. In the extreme case of four RF cavities failing, the 6D emittance increases by about 22\% and the transmission decreases by 6\%.
\section{Conclusion}\label{section_6}
In this paper, a general method is introduced for designing the rectilinear cooling channel. Using this method, two rectilinear cooling channels with separate dipole magnets before and after a bunch merging system are designed. The output emittance of the segment before bunch merging meets the requirements of the downstream system. The segment after bunch merging achieves an output transverse emittance that is half of what was achieved in previous studies, aiding in achieving a lower output emittance in the final cooling section. The cooling performance employing $\pi$-mode RF cavities is investigated. Simulations indicate no significant difference in cooling performance between $\pi$-mode and normal mode RF cavities, except that $\pi$-mode RF requires a higher gradient due to a lower transit time factor. Error analysis regarding magnetic and RF errors has been conducted for the demonstrator-like B-stage 5. Simulation results indicate that the lattice of this stage is highly robust against RF errors, including gradient, phase, position, and rotation, unless these errors reach extremely large values, which are nearly impossible in reality.  For magnetic errors, when the current, position and rotation errors are applied simultaneously, the threshold which corresponds to about 1\% reduction in transmission is (0.2\%, 0.2 mm, 0.02°). It is feasible to control the RMS value of current errors to under 0.2\% in actual practice, but controlling the RMS value of position and rotation errors to under 0.2 mm and 0.02°, respectively, might be more challenging.

For future studies, it will be interesting to investigate whether using lower frequency RF (e.g., 176 MHz) for stage 1 before bunch merging influences cooling performance. The transmission in stage 1 before bunch merging is lower compared to the other stages due to the constrained iris radius of the RF cavities. RF cavities operating at 176 MHz are larger than those at 352 MHz, potentially allowing for larger irises that may yield improved transmission. Additionally, beyond error analysis for only a demonstrator-like B-stage 5, it would be valuable to conduct a comprehensive error analysis for the two rectilinear cooling channels before and after bunch merging.
\begin{acknowledgements}
The authors would like to thank Scott Berg for his valuable feedback on the manuscript. We also appreciate the discussions with Scott Berg, Alexej Grudiev, Siara Sandra Fabbri, and other members of the International Muon Collider Collaboration. This work is supported by the China National Funds for Distinguished Young Scientists (Grant No. 12425501). Ruihu Zhu also acknowledges funding from the China Scholarship Council (File No. 202304910408). 
\end{acknowledgements}
\appendix
\section{\label{simulation_files}Simulation files}
The beam input files, Ecalc9f control files, lattice files and beam emittance calculations for all stages in the pre-merging and post-merging sections presented in Section \ref{section_4} are available at \href{https://github.com/MuonCollider-WG4/rectilinear/tree/main/2024-8-12_lattice_files}{https://github.com/MuonCollider-WG4/rectilinear/tree/main/2024-8-12\_lattice\_files}.
\section{\label{current_density}Selection of conductor material for solenoid coils }
REBCO at 20 K has been selected as the conductor material for all coils in the rectilinear cooling channel discussed in this paper. This choice allows for higher magnetic fields at potentially lower costs, and a detailed engineering design is currently being investigated by the magnet design group of the International Muon Collider Collaboration \cite{interim_report}. Fig.~(\ref{fig_current_density}) presents the engineering critical current density alongside the current density utilized in the design described in Section \ref{section_4}, plotted against the radial magnetic field. As shown in Fig.~(\ref{fig_current_density}), the current density employed in the coils throughout all stages of the design remains below the engineering critical values.
\begin{figure}[htbp]
  \centering
  \includegraphics[width=\columnwidth]{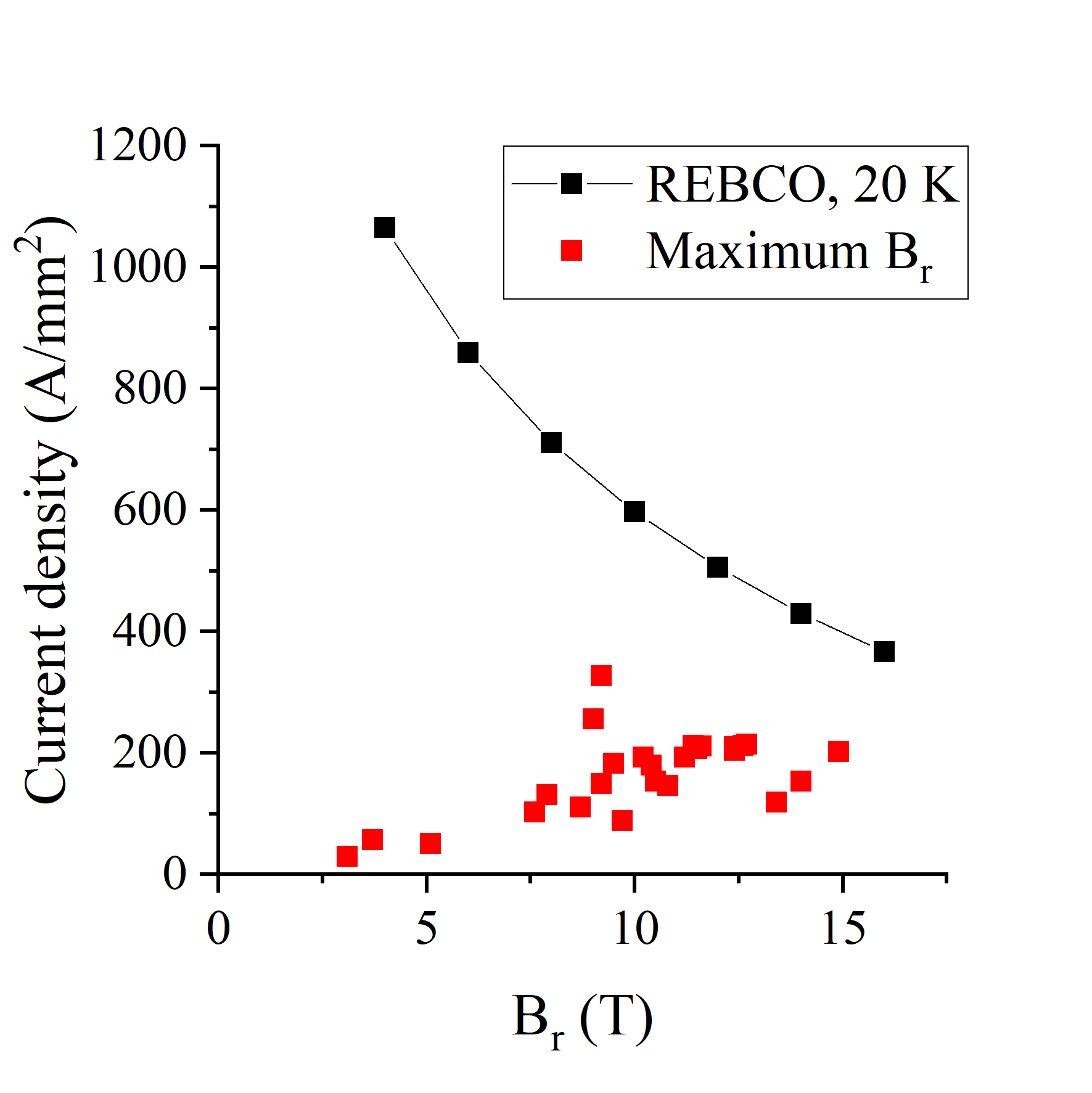}
  \caption{Current density versus radial magnetic field. Black dots: engineering critical current density using REBCO as conductor material at 20 K; Red dots: current density for all coils in the cooling channel discussed in Section \ref{section_4}.}
  \label{fig_current_density}
\end{figure}

\bibliography{apssamp}

\end{document}